\documentclass[letter,longauth]{aa} 
\usepackage{graphicx}
\usepackage{comment}
\usepackage{newtxtext,newtxmath}
 \usepackage{multirow}
 \usepackage{xcolor}
 \usepackage{dblfloatfix}
\usepackage{lineno}
\usepackage{url}
\usepackage{subcaption}
\usepackage{hyperref}

\begin{document}

\title{Discovery of EP~J175257.3--351923 as a Candidate Black Hole Low-Mass X-ray Binary}

\titlerunning{EP250916a}
   \authorrunning{Huang et al.}

\author{G.~L.~Huang\inst{1,2} \and Q.~C.~Zhao\inst{1,2} \and L.~Tao\inst{1}\fnmsep\thanks{E-mail: taolian@ihep.ac.cn} \and A.~Coleiro\inst{3} \and A.~Rau\inst{4} \and S.~Brennan\inst{4} \and C~.Y.~Dai\inst{5,6} \and R.~Soria\inst{7,8} \and F.~Cangemi\inst{3} \and  F.~Coti Zelati\inst{9,10} \and A.~Marino\inst{9,10} \and L.~Zhang\inst{1} \and S.~Guillot\inst{11} \and H.~Q.~Cheng\inst{12} \and H.~Feng\inst{1} \and D.~Götz\inst{13} \and Y.~Huang\inst{1} \and Y.~F.~Huang\inst{5,6} \and D.~Y.~Li\inst{12} \and Z.~S.~Li\inst{14} \and P.~Maggi\inst{15} \and R.~C.~Ma\inst{16} \and X.~Ma\inst{1} \and H.~W.~Pan\inst{12} \and N.~Rea\inst{9,10} \and J.~Wang\inst{12} \and Q.~Y.~Wu\inst{12} \and L.~P.~Xin\inst{12} \and W.~M.~Yuan\inst{12} \and Z.~H.~Yao\inst{12} \and G.~B.~Zhang\inst{17} \and W.~D.~Zhang\inst{12} \and S.~N.~Zhang\inst{1}
}

\institute{
Key Laboratory of Particle Astrophysics, Institute of High Energy Physics, Chinese Academy of Sciences, Beijing 100049, People’s Republic of China
\and 
University of Chinese Academy of Sciences, Chinese Academy of Sciences Beijing 100049, China
\and 
Université Paris Cité, CNRS, Astroparticule et Cosmologie, F-75013 Paris, France
\and
Max Planck Institute for Extraterrestrial Physics, Garching 85748, Germany
\and
School of Astronomy and Space Science, Nanjing University, Nanjing 210023,  China  
\and
Key Laboratory of Modern Astronomy and Astrophysics (Nanjing University), Ministry of Education, Nanjing 210023, China
\and
INAF-Osservatorio Astrofisico di Torino, Strada Osservatorio 20, I-10025 Pino Torinese, Italy  
\and
Sydney Institute for Astronomy, School of Physics A28, The University of Sydney, NSW 2006, Australia
\and
Institute of Space Sciences (ICE, CSIC), Campus UAB,
Carrer de Can Magrans s/n, E-08193 Barcelona, Spain
\and
Institut d'Estudis Espacials de Catalunya (IEEC), 08860
Castelldefels (Barcelona), Spain 
\and
IRAP, Universit\'{e} de Toulouse/OMP, CNRS, CNES, 9 avenue du Colonel Roche, BP 44346, F-31028 Toulouse Cedex 4, France 
\and
National Astronomical Observatories, Chinese Academy of Sciences, Beijing 100101, China
\and
CEA Paris-Saclay, Institut de Recherche sur les lois Fondamentales de l'Univers, 9111 Gif-sur-Yvette, France.
\and 
School of Science, Qingdao University of Technology, Qingdao 266525, China
\and
Observatoire Astronomique de Strasbourg, Université de Strasbourg, CNRS, 11 rue de l’Université, F-67000 Strasbourg, France 
\and
School of Physics and Astronomy, University of Southampton, Highfield, Southampton, SO17 1BJ, UK 
\and
Yunnan Observatory, Chinese Academy of Sciences, Kunming 650011, China 
}

\abstract{
We report the discovery of a new X-ray transient, EP~J175257.3--351923 (EP250916a), by the Einstein Probe (EP) near the Galactic plane. The outburst lasted for at least 250\,days, reached a peak 2--10 keV flux of $\sim 4 \times 10^{-10}$~erg~cm$^{-2}$~s$^{-1}$, and exhibited a fast-rise exponential-decay (FRED) profile typical of X-ray binary outbursts. The source remained in the hard state throughout the outburst, with only modest variations in the photon index ($\sim 1.6$--$2.2$) and no evidence for a spectral state transition. Broadband spectral modeling suggests a truncated disk, a weak reflection component, and a high-energy cutoff at $\sim 217$~keV, consistent with hard-state accretion in black-hole systems. No reliable optical or radio counterpart is detected within the Swift/XRT error circle. The inferred X-ray-to-optical and X-ray-to-radio flux ratios are consistent with a low-mass X-ray binary classification. Neither pulsations nor significant aperiodic variability are detected. Although the compact object cannot yet be firmly identified, the timing, spectral, and multiwavelength evidence favors EP~J175257.3--351923 as a black-hole low-mass X-ray binary candidate, highlighting EP's potential to uncover a faint, previously hidden population of X-ray binaries.
}

\keywords{X-ray transient; X-ray binary; Black hole; EP~J175257.3--351923}

\maketitle
\nolinenumbers
%

\section{Introduction}
\label{Introduction}

X-ray binaries (XRBs) are binary systems composed of a compact object, either a neutron star (NS) or a black hole (BH), and a stellar companion. 
Most low-mass XRBs (LMXBs) are transient sources that remain predominantly in a quiescent state~\citep{annurev:/content/journals/10.1146/annurev.astro.34.1.607,2016Tetarenko}. Episodically, these systems undergo X-ray outbursts lasting from weeks to months, likely driven by thermal-instability \citep{2001NewAR..45..449L}, typically reaching peak luminosities of $L_{\rm X} \sim 10^{37}$--$10^{39}$\,erg\,s$^{-1}$ \citep{2015Yan,2016Tetarenko,2023Bahramian}. The discovery of these transients has relied on a series of wide-field X-ray monitors launched since the 1970s, including Ariel-V, Granat/WATCH, RXTE/ASM, INTEGRAL, Swift/BAT, and MAXI \citep[e.g.,][]{Villa1976, Brandt1990, LevineRXTE1996, Winkler2003, BAT2005, MAXI2009}. However, such monitors often miss faint XRBs (peak luminosity $L_{\rm X} \sim 10^{36-37}\,\mathrm{erg\,s^{-1}}$) and the extreme cases---very faint XRBs (VFXBs; $L_{\rm X} \sim 10^{34-36}\,\mathrm{erg\,s^{-1}}$) \citep{Wijnands_2006}. The low luminosities of these systems most likely arise from low mass-transfer rates, due to their small accretion disks, evaporation of the inner disk into a radiatively inefficient flow, or propeller effects in the case of NS systems \citep{Bahramian2022}. The launch of the Einstein Probe (EP; \citealt{2022hxga.book...86Y,2025SCPMA..6839501Y}) greatly advances the sensitivity of wide-field monitoring, allowing the detection of previously elusive faint sources, and thus providing insights into the low-luminosity accretion regime that has rarely been explored before. 

Over its first two years of operation, EP has discovered  more than a dozen faint XRB candidates, such as EP240809a (MAXI~J1752--457; \citealt{2024ATel16765....1L} and Cheng et al. in prep.), EP240904a (EP~J182730.0--095633; \citealt{2025ApJ...991L..41C}), EP250315b (EP~J163933.2--411414; \citealt{2025ATel17083....1L} and \citealt{EPJ163933}), EP J174942.2--384834 \citep{2026ApJ..1003..224C}, and EP250623a (EP J171159.4--333253; \citealt{2026ApJ..1003...22W} and Yang et al. in prep.). These detections demonstrate EP's unique capability to explore the previously unexplored population of faint XRBs.

Here we report the discovery of a new faint XRB candidate, EP~J175257.3--351923 (designated EP250916a), and present results from the follow-up observations. This Letter is organized as follows. Sect.~\ref{Observation} describes the observations and data reduction. Sect.~\ref{Analysis} presents the analysis and results, followed by the discussion and conclusions in Sect.~\ref{Conclusions}.

\section{Observations}
\label{Observation}

EP~J175257.3--351923 was first detected by the Wide-field X-ray Telescope (WXT) on board EP at 03:33:13~UTC on 2025 September 16 \citep{2025GCN.41841....1W,2025ATel17395....1D} (Fig.~\ref{src_image}). EP/WXT continued monitoring this source until 2025 October 17, with typical exposures of $\sim$1--3~ks per epoch. Following a several-month interruption due to Sun-avoidance constraints, monitoring resumed on 2026 March 15.
Multiwavelength follow-up observations (Tab.~\ref{tab:obs_info}) were performed with the EP/Follow-up X-ray Telescope (FXT; \citealt{2021SPIE11444E..5BC}), Space-based multi-band astronomical Variable Objects Monitor (SVOM; \citealt{2016arXiv161006892W,2022svom}), Neil Gehrels Swift Observatory (Swift; \citealt{2004ApJ...611.1005G}), 
Nuclear Spectroscopic Telescope Array (NuSTAR; \citealt{2013ApJ...770..103H}) and Gamma-Ray burst Optical Near-infrared Detector (GROND; \citealt{2008PASP..120..405G}). The source region was also covered by the Very Large Array Sky Survey (VLASS; \citealt{2020PASP..132c5001L}) on 2026 February 1.

The observation with EP/FXT ($\sim$32\,h post-discovery; $\sim$3.1\,ks) detected a bright X-ray source within the EP/WXT error circle at R.A. (J2000) = $17^{\rm h}52^{\rm m}57.1^{\rm s}$ and Decl. (J2000) = $-35^{\circ}19'19.9''$, with a 10\arcsec\ uncertainty (90\% confidence; \citealt{2025GCN.41861....1D}).
Swift/XRT further refined the position to R.A. (J2000) = $17^{\rm h}52^{\rm m}57.3^{\rm s}$ and Decl. (J2000) = $-35^{\circ}19'22.9''$ ($l=355.3090^{\circ}$, $b=-4.6086^{\circ}$), with a 2.2\arcsec\ uncertainty at the 90\% confidence \citep{2025ATel17397....1I}.
We adopt the XRT position to search for the multiwavelength counterpart. Details of the data reduction are provided in Appendix~\ref{data_reduction}.

\section{Analysis and results}
\label{Analysis}
\subsection{Multiwavelength counterpart}
\label{Opt}

\begin{figure}[htbp]
    \centering
    \includegraphics[width=0.4\textwidth]{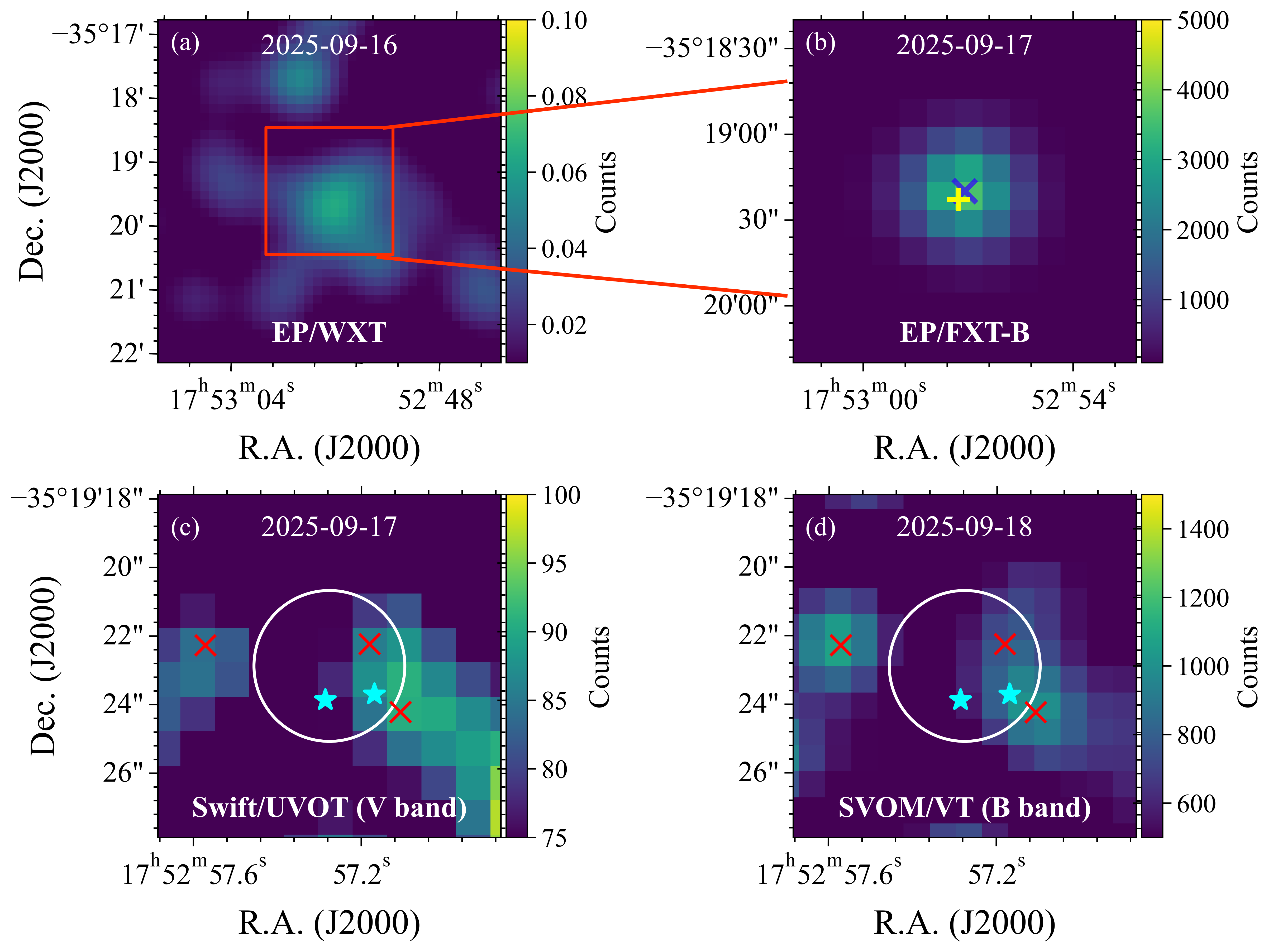}  
    \caption{Multi-wavelength images of EP~J175257.3--351923. (a) EP/WXT discovery image of the new transient. (b) EP/FXT-B follow-up image; the blue cross and yellow plus mark the EP/FXT and Swift/XRT positions, respectively. (c) Swift/UVOT $V$-band image; the white circle shows the Swift/XRT positional uncertainty (2.2\arcsec, 90\%), red crosses mark three cataloged sources detected in both Gaia and VIRAC2/VVVX, and cyan stars indicate two additional sources listed only in VIRAC2/VVVX. (d) SVOM/VT $B$-band image, with the same symbols as in panel (c).
    All images have been smoothed for display purposes.}
    \vspace{-0.3cm}
    \label{src_image}
\end{figure}

As shown in the lower panels of Fig.~\ref{src_image} and in Fig.~\ref{image_grond}, five catalogued optical/infrared sources are located within or near the Swift/XRT error circle. Three of these are recorded in both Gaia DR3 \citep{2023A&A...674A...1G} and VIRAC2/VVVX \citep{2025MNRAS.536.3707S}, while the remaining two are only catalogued in VIRAC2/VVVX. 
Photometry in this crowded field is challenging due to source confusion. We performed small-aperture photometry on the three Gaia sources using Swift/UVOT data obtained during the outburst (see Appendix~\ref{data_reduction} for details). 
Although suboptimal, this approach allows us to examine whether the optical emission traces the X-ray variability.
In contrast to the order-of-magnitude variability observed in X-rays, the optical fluxes of these sources vary by less than a factor during the outburst and show no evolutionary track comparable to that of the X-ray light curve.
For the GROND data, reliable photometry is not feasible due to crowding. These sources are therefore likely chance alignments with the EP source, although a true counterpart cannot be fully excluded.

To constrain the optical emission from the EP source, we instead derive upper limits at the XRT position (see Appendix~\ref{data_reduction}). Aperture photometry performed at the Swift/XRT error circle yields $3\sigma$ upper limits of $V > 17.77$~mag from Swift/UVOT (ObsID: 03000100002) on September 17 and $B > 18.71$~mag from SVOM/VT (ObsID: 06333) on September 18. We also searched the VLASS quick-look image without detecting a radio counterpart within the XRT error circle, deriving a $3\sigma$ upper limit of $4.44\times10^{-4}\,\mathrm{Jy\,beam^{-1}}$ at 3\,GHz (Appendix~\ref{VLASS}).

\subsection{X-ray spectral analysis}

The X-ray spectral analysis was performed with XSPEC v12.14.0b \citep{arnaud96} using data from EP/WXT (0.5--4.0\,keV), EP/FXT (0.5--10\,keV), Swift/XRT (0.3--10\,keV), SVOM/MXT (0.4--10\,keV), and NuSTAR (4--78\,keV). Interstellar absorption was modeled with tbabs adopting the solar abundances of \citet{2000ApJ...542..914W}, and a cross-calibration constant was included in joint fits. Unabsorbed X-ray fluxes were estimated using the cflux convolution model. 
The details of the spectral fitting are present in Appendix~\ref{spec_details}.

\subsection{Long-term X-ray evolution}
\begin{figure}[htbp]
    \centering
    \includegraphics[width=0.5\textwidth]{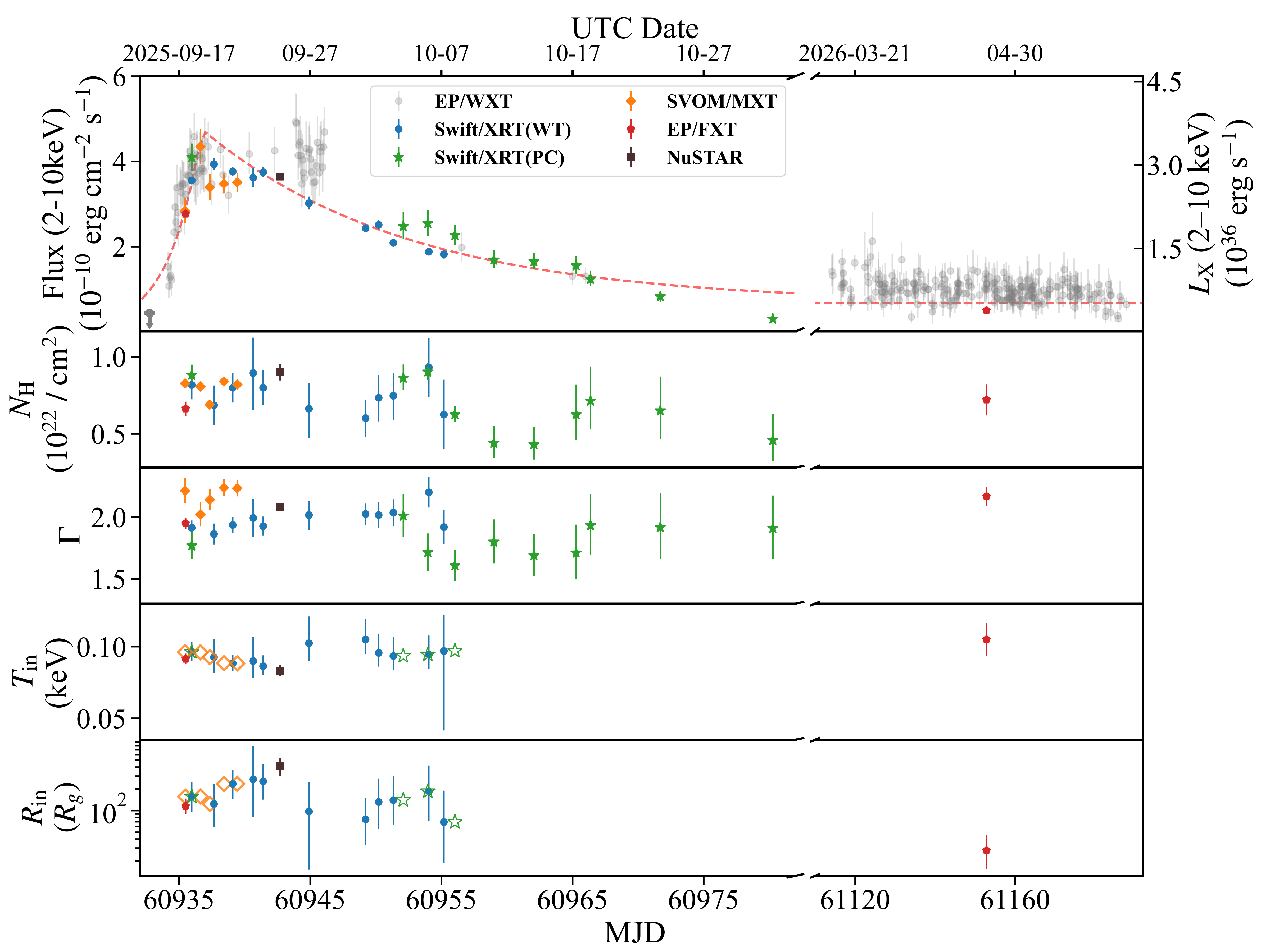}
    \caption{
    Long-term X-ray evolution: unabsorbed 2--10~keV flux/luminosity, hydrogen column density ($N_{\rm H}$), power-law index ($\Gamma$), inner disk temperature ($T_{\rm in}$), and inner disk radius ($R_{\rm in}$). Open symbols indicate fixed diskbb values. Different symbols correspond to different instruments. The red solid line represents the best-fitting FRED model (excluding the pre-outburst EP/WXT upper limit). Luminosity and $R_{\rm in}$ are calculated assuming $D = 8$~kpc, $M = 10~M_\odot$, and $i = 58\degr$ (Table~\ref{tab_fit_spec}).}
    \vspace{-0.3cm}
    \label{fig:long_lc}
\end{figure}

\begin{figure}
    \centering
    \includegraphics[width=0.33\textwidth]{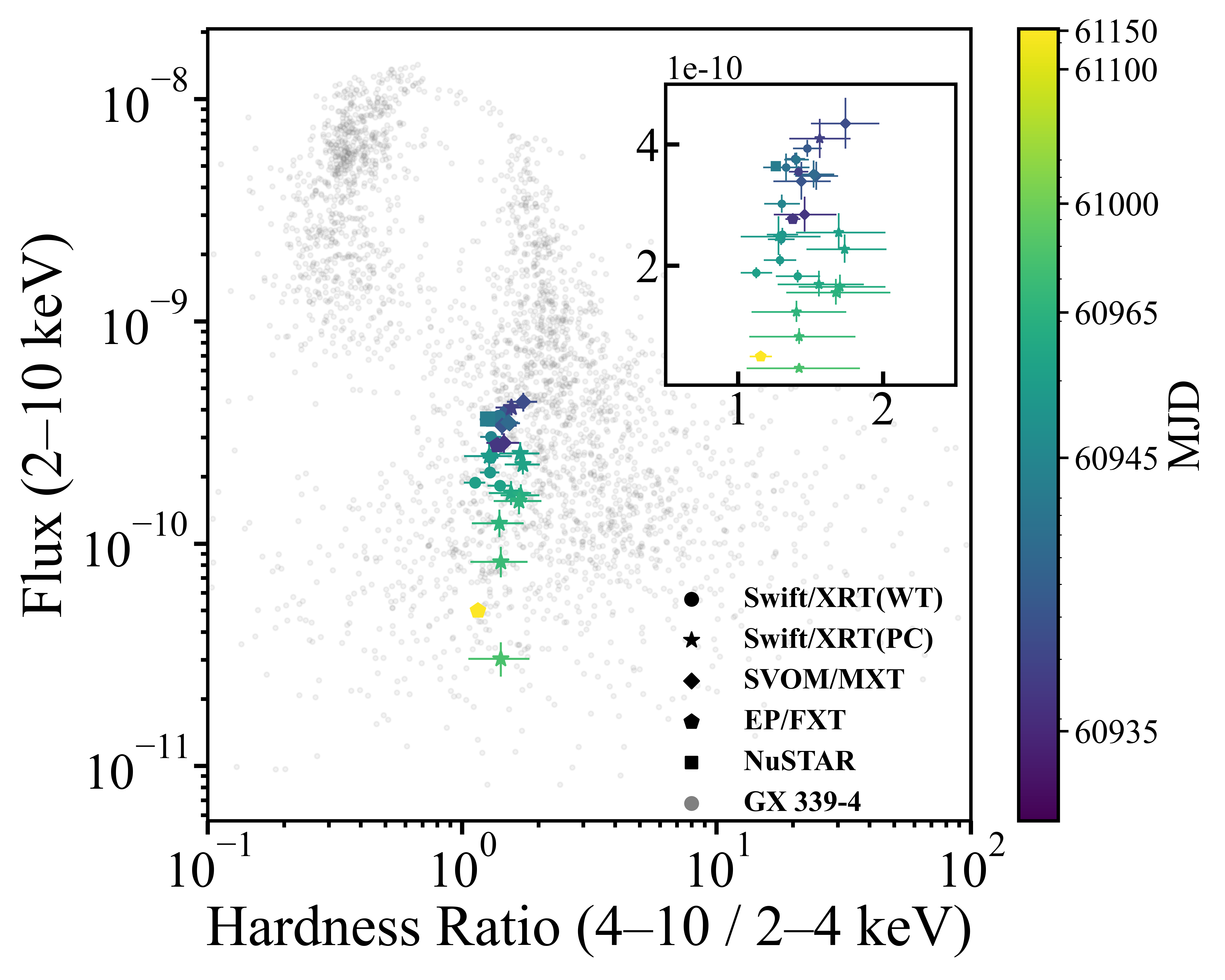}
    \caption{HID during the outburst. EP/WXT data are excluded due to the narrow energy band. The inset shows a zoomed-in view. Faint gray points show long-term MAXI data of GX~339--4 (`q'-like tracks of ``successful'' outbursts are evident), for comparison.}
    \vspace{-0.3cm}
    \label{fig:hid}
\end{figure}

The long-term evolution of the flux is shown in Fig.~\ref{fig:long_lc}. 
Archival EP/WXT observations performed on 2025 September 14 and 15 yield a pre-outburst unabsorbed $2$--$10$~keV flux upper limit of $\sim 4\times 10^{-11}\ {\rm erg\ cm^{-2}\ s^{-1}}$.
Following outburst onset at MJD~60934 (2025 September 16), the source exhibited a rapid flux increase, peaking at $\sim$MJD~60936 (2025 September 18), after which the emission entered a gradual decay phase. The X-ray light curve is well described by a fast-rise exponential-decay (FRED) profile, characterized by a rise timescale of $\sim2.5$~days and a decay timescale of $\sim15.7$~days (see Appendix$~\ref{fred}$).

$\Gamma$ remained within $\sim$1.6--2.2 throughout the outburst, with no clear evidence of spectral softening. To further investigate the state evolution during the outburst, we constructed a hardness–intensity diagram (HID; Fig.~\ref{fig:hid}). Hardness is defined as the ratio of unabsorbed fluxes in the 4--10~keV and 2--4~keV bands, and intensity as the unabsorbed flux in the 2--10~keV band. The HID indicates that the source remained in the hard state, with no evidence for a state transition.

$N_{\rm H}$ ranges from $4 \times 10^{21}$ to $9 \times 10^{21}\,\mathrm{cm^{-2}}$ during the outburst, showing no significant evolution within the uncertainties. The $N_{\rm H}$ values exceed the Galactic H\textsc{i} column density of $\sim 2 \times 10^{21}\,\mathrm{cm^{-2}}$ from the HI4PI survey \citep{2016A&A...594A.116H}\footnote{\url{https://heasarc.gsfc.nasa.gov/cgi-bin/Tools/w3nh/w3nh.pl}}, hinting at potential intrinsic absorption. However, given the uncertainties in the HI4PI Galactic estimate, this apparent excess may instead stem from differential reddening in the Galactic bulge.

\subsection{Timing analysis}

We inspected all available X-ray light curves and found no evidence of Type-I bursts. Average power spectral densities (PSDs) were computed using stingray v2.3.2 \citep{2019ApJ...881...39H,Bachetti2024}, based on 512~s segments from 0.005\,s barycenter-corrected light curves of EP/FXT-B, Swift/XRT (WT mode), SVOM/MXT, and NuSTAR. No significant coherent pulsations, quasi-periodic oscillations (QPOs), or red noise were detected, whereas the higher signal-to-noise XMM-Newton observation revealed a weak QPO at a frequency of $\sim$13~Hz \citep{2026arXiv260608752J}. The averaged power spectra (fractional-rms normalization; Fig.~\ref{pds_nustar}) from NuSTAR gave a 99\% upper limit of 10.3\% on the integrated rms over 0.1--64~Hz. An acceleration search using PRESTO \citep{Ransom2011} found no significant pulsations, implying a 99\% upper limit of $\sim6$--7\% on the pulsed fraction.

\section{Discussion and conclusions}
\label{Conclusions}

We report the discovery of a new long-duration X-ray transient, EP~J175257.3--351923, located at Galactic latitude $b=-4.6086^{\circ}$, close to the Galactic plane where the space density of XRBs reaches its maximum \citep{Grimm_2002,2023Bahramian}. 
The source exhibits characteristics analogous to those of XRBs. Its long-term X-ray light curve exhibits a characteristic FRED morphology, a hallmark of transient XRB outbursts \citep{Chen_FRED,Remillard_review}. The outburst persists for at least $\sim$250 days, well within the typical duration range observed in Galactic XRB transients \citep{2016Tetarenko,2025ApJS..279...57H}. The source maintains remarkable spectral stability throughout the outburst, remaining in the hard state with only limited variation in photon index. Such behavior is commonly referred to as a ``failed outburst'' and has been reported in hard-state outbursts of XRBs \citep{2016Tetarenko,2015Yan}. 
Furthermore, broadband spectral modeling suggests a truncated accretion disk together with a reflection component, with a photon index $\Gamma \sim 1.6$--2.2 and a high-energy cutoff at $\sim217$~keV, broadly consistent with the canonical hard-state characteristics of XRBs \citep{2007A&ARv..15....1D,2023hxga.book..139B}.

No reliable optical or radio counterpart was detected within the Swift/XRT localization region in the available Swift/UVOT, SVOM/VT, GROND and VLASS images. The optical non-detection is likely due to the combined effects of strong interstellar extinction, weak optical brightening associated with the faint X-ray outburst, and source confusion in the crowded Galactic-plane environment. The radio non-detection may also reflects weak emission from the faint outburst.
Assuming a representative distance of 8~kpc and a reddening of $E(B-V)=0.40$~mag \citep{2025ApJS..280...15W}, we derive a $3\sigma$ upper limit on the dereddened $B$-band magnitude, $B_0 \gtrsim 18.58$~mag (using the extinction law of \citealt{1989ApJ...345..245C}). Using quasi-simultaneous X-ray and optical data, we derive a lower limit on the X-ray-to-optical flux ratio, defined as $\xi = B_0 + 2.5\log F_X$, yielding $\xi \gtrsim 21.75$. This is consistent with values typically observed in LMXBs \citep{1995xrbi.nasa.....L}. Furthermore, assuming a distance of 8~kpc, both the X-ray-to-optical and X-ray-to-radio luminosity ratios broadly agree with the hard-state LMXB correlations (see Figs.~\ref{fig_lv_lx} and ~\ref{fig_lrlx}; \citealt{2006MNRAS.371.1334R,arash_bahramian_2022_7059313}).
Collectively, the temporal, spectral, and multiwavelength properties favor this source as a LMXB candidate, consistent with \citet{2026arXiv260608752J}.

EP~J175257.3--351923, with an X-ray luminosity of the order of $\sim10^{36}\,\mathrm{erg\,s^{-1}}$, can be classified as a faint or very faint XRB candidate according to \citet{Wijnands_2006}. In particular, in many VFXBs the companions are expected to be intrinsically faint (e.g., low-mass main-sequence stars). The multiwavelength correlations observed for EP~J175257.3--351923 are consistent with this scenario. Moreover, the long-term light-curve morphology, and photon-index evolution of this EP source resemble those of VFXBs such as EP~J182730.0$-$095633 \citep{2025ApJ...991L..41C} and IGR~J17285$-$2922 \citep{Stoop2021}, both of which have been suggested as black hole candidates. Interestingly, a stable QPO at $\sim40$~mHz with a fractional rms amplitude of $\sim20$--30\,\% has been detected in EP~J182730.0$-$095633. In contrast, we do not detect any significant QPO in EP~J175257.3--351923; the PSD is largely featureless. Nevertheless, featureless PSDs have also been reported in several VFXBs \citep{2014MNRAS.442..372K,2018IGRJ17062-6143,2022MNRAS.515.3838M}. Furthermore, the presence of a truncated accretion disk and a reflection component is in good agreement with the properties reported for the VFXBs IGR~J17062--6143 and MAXI~J1848--015 \citep{2018IGRJ17062-6143,2022ApJ...927..190P}.

Based on the FRED modeling, the decay timescale of $\sim15.7$~days implies an outer disk radius of $\sim10^{4}\,R_{\mathrm{g}}$ (Appendix~\ref{binary}). Assuming this radius is comparable to the circularization radius, we derive an estimated orbital period of $\sim 4.3$~hr. Although these estimates are subject to considerable uncertainties arising from both the model assumptions and the choice of system parameters, this source is likely a short-period system with a relatively small accretion disk. An analogous example is Swift~J1357.2$-$0933, a short-period ($P \approx 3$~hr) black hole XRB transient in the VFXB regime \citep{2015MNRAS.454.2199M}. Such short-period systems represent a proposed channel for the formation of VFXBs \citep{2015MNRAS.447.3034H}.

Regarding the nature of the compact object, the absence of coherent pulsations, Type-I X-ray bursts, and dynamical mass measurements leaves both NS and BH interpretations viable. However, the high cut-off energy of $\sim217$~keV inferred from spectral modeling favors a BH system \citep{2017MNRAS.466..194B}. This is because seed photons from an NS surface and boundary layer enhance Compton cooling, generally leading to lower cut-off energies, whereas the absence of a solid surface in BHs allows deeper gravitational energy release and higher electron temperatures \citep{2017MNRAS.466..194B,2024Univ...10..446P}.

Besides an XRB, alternative interpretations such as a classical nova or a tidal disruption event (TDE) appear less likely. Classical novae are typically associated with prominent supersoft X-ray emission \citep{2021ARA&A..59..391C}, which is absent in this source. Thermal TDEs generally exhibit soft X-ray spectra \citep{2023MNRAS.519.5828M}, whereas the X-ray spectrum of this EP source remains persistently hard. Moreover, unlike jetted TDEs such as Swift J1644+57 \citep{2015MNRAS.450.2824M} and AT2022cmc \citep{2023MNRAS.521..389R}, which display bright radio emission detectable for months to years, no radio counterpart is found in the VLASS images at the position of this source, further disfavouring the TDE interpretation.

Overall, future multiwavelength observations, particularly with large-aperture optical telescopes, will be essential to further constrain the nature of the compact object and to fully characterize the physical properties of this system. Nevertheless, this source provides an interesting case for understanding weak outbursts and demonstrates EP's capability to discover faint XRBs.

\begin{acknowledgements}
We thank the anonymous referee for comments that helped improve the paper. This work made use of data from EP, SVOM, NuSTAR, and Swift. This work was supported by the National Key R\&D Program of China (Nos.~2025YFF0511102 and 2021YFA0718500), the National Natural Science Foundation of China (Nos.~12122306, 12025301, and 12103027), the Strategic Priority Research Program of the Chinese Academy of Sciences, and China's Space Origins Exploration Program. SG acknowledges the support of the CNES.
F.C.Z. is supported by the Ram\'on y Cajal fellowship (RYC2021-030888-I), Spanish grant ID2023-153099NA-I00, and the program Unidad de Excelencia Maria de Maeztu CEX2020-001058-M. 
Part of the funding for GROND (both hardware and personnel) was generously granted by the Leibniz-Prize to G. Hasinger (DFG grant HA 1850/28-1) and by the Thüringer Landessternwarte Tautenburg. Y.~F.~H. also acknowledges the support from the Xinjiang Tianchi Program.

\end{acknowledgements}
\bibliographystyle{yahapj} 
\bibliography{Bibliography} 

@ARTICLE{2016A&A...594A.116H,
       author = {{HI4PI Collaboration} and {Ben Bekhti}, N. and {Fl{\"o}er}, L. and {Keller}, R. and {Kerp}, J. and {Lenz}, D. and {Winkel}, B. and {Bailin}, J. and {Calabretta}, M.~R. and {Dedes}, L. and {Ford}, H.~A. and {Gibson}, B.~K. and {Haud}, U. and {Janowiecki}, S. and {Kalberla}, P.~M.~W. and {Lockman}, F.~J. and {McClure-Griffiths}, N.~M. and {Murphy}, T. and {Nakanishi}, H. and {Pisano}, D.~J. and {Staveley-Smith}, L.},
        title = "{HI4PI: A full-sky H I survey based on EBHIS and GASS}",
      journal = {\aap},
         year = 2016,
        month = oct,
       volume = {594},
          eid = {A116},
        pages = {A116},
          doi = {10.1051/0004-6361/201629178},
archivePrefix = {arXiv},
       eprint = {1610.06175},
 primaryClass = {astro-ph.GA},
       adsurl = {https://ui.adsabs.harvard.edu/abs/2016A&A...594A.116H}
}

@ARTICLE{2000ApJ...542..914W,
       author = {{Wilms}, J. and {Allen}, A. and {McCray}, R.},
        title = "{On the Absorption of X-Rays in the Interstellar Medium}",
      journal = {\apj},
         year = 2000,
        month = oct,
       volume = {542},
       number = {2},
        pages = {914-924},
          doi = {10.1086/317016},
archivePrefix = {arXiv},
       eprint = {astro-ph/0008425},
 primaryClass = {astro-ph},
       adsurl = {https://ui.adsabs.harvard.edu/abs/2000ApJ...542..914W}
}

@article{annurev:/content/journals/10.1146/annurev.astro.34.1.607,
   author = "Tanaka, Y. and Shibazaki, N.",
   title = "X-RAY NOVAE", 
   journal= "Annual Review of Astronomy and Astrophysics",
   year = "1996",
   volume = "34",
   number = "Volume 34, 1996",
   pages = "607-644",
   doi = "https://doi.org/10.1146/annurev.astro.34.1.607",
   url = "https://www.annualreviews.org/content/journals/10.1146/annurev.astro.34.1.607",
   publisher = "Annual Reviews",
   issn = "1545-4282",
   type = "Journal Article",
}

@ARTICLE{2001NewAR..45..449L,
       author = {{Lasota}, Jean-Pierre},
        title = "{The disc instability model of dwarf novae and low-mass X-ray binary transients}",
      journal = {\nar},
         year = 2001,
        month = jun,
       volume = {45},
       number = {7},
        pages = {449-508},
          doi = {10.1016/S1387-6473(01)00112-9},
archivePrefix = {arXiv},
       eprint = {astro-ph/0102072},
 primaryClass = {astro-ph},
       adsurl = {https://ui.adsabs.harvard.edu/abs/2001NewAR..45..449L}
}

@INCOLLECTION{2022hxga.book...86Y,
       author = {{Yuan}, Weimin and {Zhang}, Chen and {Chen}, Yong and {Ling}, Zhixing},
        title = "{The Einstein Probe Mission}",
    booktitle = {Handbook of X-ray and Gamma-ray Astrophysics},
         year = 2022,
       editor = {{Bambi}, Cosimo and {Sangangelo}, Andrea},
          eid = {86},
        pages = {86},
          doi = {10.1007/978-981-16-4544-0_151-1},
       adsurl = {https://ui.adsabs.harvard.edu/abs/2022hxga.book...86Y}
}

@INPROCEEDINGS{2021SPIE11444E..5BC,
       author = {{Chen}, Yong and {Cui}, WeiWei and {Han}, DaWei and {Wang}, Juan and {Yang}, YanJi and {Wang}, YuSa and {Li}, Wei and {Ma}, Jia and {Xu}, YuPeng and {Lu}, FangJun and {Chen}, HouLei and {Tang}, QingJun and {Yuan}, Weimin and {Friedrich}, Peter and {Meidinger}, Norbert and {Keil}, Isabell and {Burwitz}, Vadim and {Eder}, Josef and {Hartmann}, Katinka and {Nandra}, Kirpal and {Keereman}, Arnoud and {Santovincenzo}, Andrea and {Vernani}, Dervis and {Bianucci}, Giovanni and {Valsecchi}, Giuseppe and {Wang}, Bo and {Wang}, LangPing and {Wang}, DianLong and {Li}, Duo and {Sheng}, LiZhi and {Qiang}, PengFei and {Shi}, RongRong and {Chao}, XiangYu and {Song}, Zeyu and {Zhang}, Ziliang and {Huo}, Jia and {Wang}, Hao and {Cong}, Min and {Yang}, XiongTao and {Hou}, Dongjie and {Zhao}, XiaoFan and {Zhao}, ZiJian and {Chen}, TianXiang and {Li}, MaoShun and {Zhang}, Tong and {Luo}, LaiDan and {Xu}, JingJing and {Li}, Gang and {Zhang}, Qian and {Bi}, XiYan and {Zhu}, YuXuan and {Yu}, Nian and {Chen}, Can and {Lv}, ZhongHua and {Lu}, Bing and {Zhang}, JiaWei},
        title = "{Status of the follow-up x-ray telescope onboard the Einstein Probe satellite}",
    booktitle = {Society of Photo-Optical Instrumentation Engineers (SPIE) Conference Series},
         year = 2021,
       editor = {{den Herder}, Jan-Willem A. and {Nikzad}, Shouleh and {Nakazawa}, Kazuhiro},
       series = {Society of Photo-Optical Instrumentation Engineers (SPIE) Conference Series},
       volume = {11444},
        month = jan,
          eid = {114445B},
        pages = {114445B},
          doi = {10.1117/12.2562311},
       adsurl = {https://ui.adsabs.harvard.edu/abs/2021SPIE11444E..5BC}
}

@ARTICLE{2025SCPMA..6839501Y,
       author = {{Yuan}, Weimin and {Dai}, Lixin and {Feng}, Hua and {Jin}, Chichuan and {Jonker}, Peter and {Kuulkers}, Erik and {Liu}, Yuan and {Nandra}, Kirpal and {O'Brien}, Paul and {Piro}, Luigi and {Rau}, Arne and {Rea}, Nanda and {Sanders}, Jeremy and {Tao}, Lian and {Wang}, Junfeng and {Wu}, Xuefeng and {Zhang}, Bing and {Zhang}, Shuangnan and {Ai}, Shunke and {Buchner}, Johannes and {Bulbul}, Esra and {Chen}, Hechao and {Chen}, Minghua and {Chen}, Yong and {Chen}, Yu-Peng and {Coleiro}, Alexis and {Coti Zelati}, Francesco and {Dai}, Zigao and {Fan}, Xilong and {Fan}, Zhou and {Friedrich}, Susanne and {Gao}, He and {Ge}, Chong and {Ge}, Mingyu and {Geng}, Jinjun and {Ghirlanda}, Giancarlo and {Gianfagna}, Giulia and {Gou}, Lijun and {Guillot}, S{\'e}bastien and {Hou}, Xian and {Hu}, Jingwei and {Huang}, Yongfeng and {Ji}, Long and {Jia}, Shumei and {Komossa}, S. and {Kong}, Albert K.~H. and {Lan}, Lin and {Li}, An and {Li}, Ang and {Li}, Chengkui and {Li}, Dongyue and {Li}, Jian and {Li}, Zhaosheng and {Ling}, Zhixing and {Liu}, Ang and {Liu}, Jinzhong and {Liu}, Liangduan and {Liu}, Zhu and {Luo}, Jiawei and {Ma}, Ruican and {Maggi}, Pierre and {Maitra}, Chandreyee and {Marino}, Alessio and {Ng}, Stephen Chi-Yung and {Pan}, Haiwu and {Rukdee}, Surangkhana and {Soria}, Roberto and {Sun}, Hui and {Tam}, Pak-Hin Thomas and {Thakur}, Aishwarya Linesh and {Tian}, Hui and {Troja}, Eleonora and {Wang}, Wei and {Wang}, Xiangyu and {Wang}, Yanan and {Wei}, Junjie and {Wen}, Sixiang and {Wu}, Jianfeng and {Wu}, Ting and {Xiao}, Di and {Xu}, Dong and {Xu}, Renxin and {Xu}, Yanjun and {Xu}, Yu and {Yang}, Haonan and {You}, Bei and {Yu}, Heng and {Yu}, Yunwei and {Zhang}, Binbin and {Zhang}, Chen and {Zhang}, Guobao and {Zhang}, Liang and {Zhang}, Wenda and {Zhang}, Yu and {Zhou}, Ping and {Zou}, Zecheng},
        title = "{Science objectives of the Einstein Probe mission}",
      journal = {Science China Physics, Mechanics, and Astronomy},
         year = 2025,
        month = mar,
       volume = {68},
       number = {3},
          eid = {239501},
        pages = {239501},
          doi = {10.1007/s11433-024-2600-3},
archivePrefix = {arXiv},
       eprint = {2501.07362},
 primaryClass = {astro-ph.HE},
       adsurl = {https://ui.adsabs.harvard.edu/abs/2025SCPMA..6839501Y}
}

@ARTICLE{2025GCN.41841....1W,
       author = {{Wu}, Q.~Y. and {Dai}, C.~Y. and {Li}, D.~Y. and {Pan}, H.~W. and {Einstein Probe Team}},
        title = "{EP250916a: Einstein Probe detection of an X-ray transient}",
      journal = {GRB Coordinates Network},
         year = 2025,
        month = sep,
       volume = {41841},
        pages = {1},
       adsurl = {https://ui.adsabs.harvard.edu/abs/2025GCN.41841....1W}
}

@ARTICLE{2025GCN.41861....1D,
       author = {{Dai}, C.~Y. and {Wu}, Q.~Y. and {Li}, D.~Y. and {Pan}, H.~W. and {Einstein Probe Team}},
        title = "{EP250916a: EP-FXT follow-up observation}",
      journal = {GRB Coordinates Network},
         year = 2025,
        month = sep,
       volume = {41861},
        pages = {1},
       adsurl = {https://ui.adsabs.harvard.edu/abs/2025GCN.41861....1D}
}

@ARTICLE{2025ATel17395....1D,
       author = {{Dai}, C.~Y. and {Wu}, Q.~Y. and {Li}, D.~Y. and {Pan}, H.~W.},
        title = "{EP250916a: Einstein Probe detection of an X-ray transient}",
      journal = {The Astronomer's Telegram},
         year = 2025,
        month = sep,
       volume = {17395},
        pages = {1},
       adsurl = {https://ui.adsabs.harvard.edu/abs/2025ATel17395....1D}
}

@ARTICLE{2004ApJ...611.1005G,
       author = {{Gehrels}, N. and {Chincarini}, G. and {Giommi}, P. and {Mason}, K.~O. and {Nousek}, J.~A. and {Wells}, A.~A. and {White}, N.~E. and {Barthelmy}, S.~D. and {Burrows}, D.~N. and {Cominsky}, L.~R. and {Hurley}, K.~C. and {Marshall}, F.~E. and {M{\'e}sz{\'a}ros}, P. and {Roming}, P.~W.~A. and {Angelini}, L. and {Barbier}, L.~M. and {Belloni}, T. and {Campana}, S. and {Caraveo}, P.~A. and {Chester}, M.~M. and {Citterio}, O. and {Cline}, T.~L. and {Cropper}, M.~S. and {Cummings}, J.~R. and {Dean}, A.~J. and {Feigelson}, E.~D. and {Fenimore}, E.~E. and {Frail}, D.~A. and {Fruchter}, A.~S. and {Garmire}, G.~P. and {Gendreau}, K. and {Ghisellini}, G. and {Greiner}, J. and {Hill}, J.~E. and {Hunsberger}, S.~D. and {Krimm}, H.~A. and {Kulkarni}, S.~R. and {Kumar}, P. and {Lebrun}, F. and {Lloyd-Ronning}, N.~M. and {Markwardt}, C.~B. and {Mattson}, B.~J. and {Mushotzky}, R.~F. and {Norris}, J.~P. and {Osborne}, J. and {Paczynski}, B. and {Palmer}, D.~M. and {Park}, H.-S. and {Parsons}, A.~M. and {Paul}, J. and {Rees}, M.~J. and {Reynolds}, C.~S. and {Rhoads}, J.~E. and {Sasseen}, T.~P. and {Schaefer}, B.~E. and {Short}, A.~T. and {Smale}, A.~P. and {Smith}, I.~A. and {Stella}, L. and {Tagliaferri}, G. and {Takahashi}, T. and {Tashiro}, M. and {Townsley}, L.~K. and {Tueller}, J. and {Turner}, M.~J.~L. and {Vietri}, M. and {Voges}, W. and {Ward}, M.~J. and {Willingale}, R. and {Zerbi}, F.~M. and {Zhang}, W.~W.},
        title = "{The Swift Gamma-Ray Burst Mission}",
      journal = {\apj},
         year = 2004,
        month = aug,
       volume = {611},
       number = {2},
        pages = {1005-1020},
          doi = {10.1086/422091},
archivePrefix = {arXiv},
       eprint = {astro-ph/0405233},
 primaryClass = {astro-ph},
       adsurl = {https://ui.adsabs.harvard.edu/abs/2004ApJ...611.1005G}
}

@ARTICLE{2013ApJ...770..103H,
       author = {{Harrison}, Fiona A. and {Craig}, William W. and {Christensen}, Finn E. and {Hailey}, Charles J. and {Zhang}, William W. and {Boggs}, Steven E. and {Stern}, Daniel and {Cook}, W. Rick and {Forster}, Karl and {Giommi}, Paolo and {Grefenstette}, Brian W. and {Kim}, Yunjin and {Kitaguchi}, Takao and {Koglin}, Jason E. and {Madsen}, Kristin K. and {Mao}, Peter H. and {Miyasaka}, Hiromasa and {Mori}, Kaya and {Perri}, Matteo and {Pivovaroff}, Michael J. and {Puccetti}, Simonetta and {Rana}, Vikram R. and {Westergaard}, Niels J. and {Willis}, Jason and {Zoglauer}, Andreas and {An}, Hongjun and {Bachetti}, Matteo and {Barri{\`e}re}, Nicolas M. and {Bellm}, Eric C. and {Bhalerao}, Varun and {Brejnholt}, Nicolai F. and {Fuerst}, Felix and {Liebe}, Carl C. and {Markwardt}, Craig B. and {Nynka}, Melania and {Vogel}, Julia K. and {Walton}, Dominic J. and {Wik}, Daniel R. and {Alexander}, David M. and {Cominsky}, Lynn R. and {Hornschemeier}, Ann E. and {Hornstrup}, Allan and {Kaspi}, Victoria M. and {Madejski}, Greg M. and {Matt}, Giorgio and {Molendi}, Silvano and {Smith}, David M. and {Tomsick}, John A. and {Ajello}, Marco and {Ballantyne}, David R. and {Balokovi{\'c}}, Mislav and {Barret}, Didier and {Bauer}, Franz E. and {Blandford}, Roger D. and {Brandt}, W. Niel and {Brenneman}, Laura W. and {Chiang}, James and {Chakrabarty}, Deepto and {Chenevez}, Jerome and {Comastri}, Andrea and {Dufour}, Francois and {Elvis}, Martin and {Fabian}, Andrew C. and {Farrah}, Duncan and {Fryer}, Chris L. and {Gotthelf}, Eric V. and {Grindlay}, Jonathan E. and {Helfand}, David J. and {Krivonos}, Roman and {Meier}, David L. and {Miller}, Jon M. and {Natalucci}, Lorenzo and {Ogle}, Patrick and {Ofek}, Eran O. and {Ptak}, Andrew and {Reynolds}, Stephen P. and {Rigby}, Jane R. and {Tagliaferri}, Gianpiero and {Thorsett}, Stephen E. and {Treister}, Ezequiel and {Urry}, C. Megan},
        title = "{The Nuclear Spectroscopic Telescope Array (NuSTAR) High-energy X-Ray Mission}",
      journal = {\apj},
         year = 2013,
        month = jun,
       volume = {770},
       number = {2},
          eid = {103},
        pages = {103},
          doi = {10.1088/0004-637X/770/2/103},
archivePrefix = {arXiv},
       eprint = {1301.7307},
 primaryClass = {astro-ph.IM},
       adsurl = {https://ui.adsabs.harvard.edu/abs/2013ApJ...770..103H}
}

@ARTICLE{2016arXiv161006892W,
       author = {{Wei}, J. and {Cordier}, B. and {Antier}, S. and {Antilogus}, P. and {Atteia}, J. -L. and {Bajat}, A. and {Basa}, S. and {Beckmann}, V. and {Bernardini}, M.~G. and {Boissier}, S. and {Bouchet}, L. and {Burwitz}, V. and {Claret}, A. and {Dai}, Z. -G. and {Daigne}, F. and {Deng}, J. and {Dornic}, D. and {Feng}, H. and {Foglizzo}, T. and {Gao}, H. and {Gehrels}, N. and {Godet}, O. and {Goldwurm}, A. and {Gonzalez}, F. and {Gosset}, L. and {G{\"o}tz}, D. and {Gouiffes}, C. and {Grise}, F. and {Gros}, A. and {Guilet}, J. and {Han}, X. and {Huang}, M. and {Huang}, Y. -F. and {Jouret}, M. and {Klotz}, A. and {La Marle}, O. and {Lachaud}, C. and {Le Floch}, E. and {Lee}, W. and {Leroy}, N. and {Li}, L. -X. and {Li}, S.~C. and {Li}, Z. and {Liang}, E. -W. and {Lyu}, H. and {Mercier}, K. and {Migliori}, G. and {Mochkovitch}, R. and {O'Brien}, P. and {Osborne}, J. and {Paul}, J. and {Perinati}, E. and {Petitjean}, P. and {Piron}, F. and {Qiu}, Y. and {Rau}, A. and {Rodriguez}, J. and {Schanne}, S. and {Tanvir}, N. and {Vangioni}, E. and {Vergani}, S. and {Wang}, F. -Y. and {Wang}, J. and {Wang}, X. -G. and {Wang}, X. -Y. and {Watson}, A. and {Webb}, N. and {Wei}, J.~J. and {Willingale}, R. and {Wu}, C. and {Wu}, X. -F. and {Xin}, L. -P. and {Xu}, D. and {Yu}, S. and {Yu}, W. -F. and {Yu}, Y. -W. and {Zhang}, B. and {Zhang}, S. -N. and {Zhang}, Y. and {Zhou}, X.~L.},
        title = "{The Deep and Transient Universe in the SVOM Era: New Challenges and Opportunities - Scientific prospects of the SVOM mission}",
      journal = {arXiv e-prints},
         year = 2016,
        month = oct,
          eid = {arXiv:1610.06892},
        pages = {arXiv:1610.06892},
          doi = {10.48550/arXiv.1610.06892},
archivePrefix = {arXiv},
       eprint = {1610.06892},
 primaryClass = {astro-ph.IM},
       adsurl = {https://ui.adsabs.harvard.edu/abs/2016arXiv161006892W}
}

@ARTICLE{2025ApJ...991L..41C,
       author = {{Cheng}, H.~Q. and {Zhao}, Q.~C. and {Tao}, L. and {Feng}, H. and {Coti Zelati}, F. and {Pan}, H.~W. and {Wang}, A.~L. and {Wang}, Y.~N. and {Ge}, M.~Y. and {Rau}, A. and {Marino}, A. and {Zhang}, L. and {Zhang}, W.~J. and {Carotenuto}, F. and {Ji}, L. and {Jin}, C.~C. and {Li}, D.~Y. and {Liu}, B.~F. and {Liu}, Y. and {Qiao}, E.~L. and {Rea}, N. and {Soria}, R. and {Wang}, S. and {Yan}, Z. and {Yuan}, W. and {Zhang}, B. and {Zhang}, G.~B. and {Zhang}, S.~N. and {Zhang}, W.~D. and {Beardmore}, A. and {Bright}, J.~S. and {Chen}, X.~L. and {Fan}, Z. and {Fu}, S.~Y. and {Fynbo}, J.~P.~U. and {Hu}, J.~W. and {Jin}, J.~J. and {Jonker}, P.~G. and {Kong}, A.~K.~H. and {Kuulkers}, E. and {Li}, C.~K. and {Li}, H.~L. and {Lin}, Z.~K. and {Liu}, C.~X. and {Liu}, H.-Y. and {Liu}, J.~Z. and {Liu}, X.~W. and {Lu}, Z. and {Maitra}, C. and {Mu}, H.~Y. and {Ng}, C.-Y. and {Qiu}, Y.~L. and {Tinyanont}, S. and {Wang}, Y. and {Wen}, S.~X. and {Weng}, S.~S. and {Wu}, Jianfeng and {Xu}, D. and {Yan}, Y.~K. and {Yang}, Y.-P. and {Zhang}, P. and {Zhang}, S. and {Zhao}, Q. and {Cai}, Z.~M. and {Chen}, Y. and {Chen}, Y.~F. and {Cui}, C.~Z. and {Cui}, W.~W. and {Hu}, H.~B. and {Huang}, M.~H. and {Jia}, S.~M. and {Jin}, G. and {Ling}, Z.~X. and {Liu}, H.~Q. and {Sun}, S.~L. and {Sun}, X.~J. and {Xu}, Y.~F. and {Zhang}, C. and {Zhang}, M. and {Zhang}, Y.~H.},
        title = "{Einstein Probe Discovery of EP J182730.0‑095633: A New Black Hole X-Ray Binary Candidate in Faint Outburst?}",
      journal = {\apjl},
         year = 2025,
        month = oct,
       volume = {991},
       number = {2},
          eid = {L41},
        pages = {L41},
          doi = {10.3847/2041-8213/adf104},
archivePrefix = {arXiv},
       eprint = {2507.12876},
 primaryClass = {astro-ph.HE},
       adsurl = {https://ui.adsabs.harvard.edu/abs/2025ApJ...991L..41C}
}

@ARTICLE{2024Univ...10..446P,
       author = {{Pszota}, G{\'a}bor},
        title = "{High-Energy Spectra of Black Hole and Neutron Star Low-Mass X-Ray Binaries}",
      journal = {Universe},
         year = 2024,
        month = dec,
       volume = {10},
       number = {12},
          eid = {446},
        pages = {446},
          doi = {10.3390/universe10120446},
       adsurl = {https://ui.adsabs.harvard.edu/abs/2024Univ...10..446P}
}

@ARTICLE{2016Tetarenko,
       author = {{Tetarenko}, B.~E. and {Sivakoff}, G.~R. and {Heinke}, C.~O. and {Gladstone}, J.~C.},
        title = "{WATCHDOG: A Comprehensive All-sky Database of Galactic Black Hole X-ray Binaries}",
      journal = {\apjs},
         year = 2016,
        month = feb,
       volume = {222},
       number = {2},
          eid = {15},
        pages = {15},
          doi = {10.3847/0067-0049/222/2/15},
archivePrefix = {arXiv},
       eprint = {1512.00778},
 primaryClass = {astro-ph.HE},
       adsurl = {https://ui.adsabs.harvard.edu/abs/2016ApJS..222...15T}
}

@INCOLLECTION{2023Bahramian,
       author = {{Bahramian}, Arash and {Degenaar}, Nathalie},
        title = "{Low-Mass X-ray Binaries}",
    booktitle = {Handbook of X-ray and Gamma-ray Astrophysics},
         year = 2023,
          eid = {120},
        pages = {120},
          doi = {10.1007/978-981-16-4544-0_94-1},
       adsurl = {https://ui.adsabs.harvard.edu/abs/2023hxga.book..120B}
}

@ARTICLE{2015Yan,
       author = {{Yan}, Zhen and {Yu}, Wenfei},
        title = "{X-Ray Outbursts of Low-mass X-Ray Binary Transients Observed in the RXTE Era}",
      journal = {\apj},
         year = 2015,
        month = jun,
       volume = {805},
       number = {2},
          eid = {87},
        pages = {87},
          doi = {10.1088/0004-637X/805/2/87},
archivePrefix = {arXiv},
       eprint = {1408.5146},
 primaryClass = {astro-ph.HE},
       adsurl = {https://ui.adsabs.harvard.edu/abs/2015ApJ...805...87Y}
}

@ARTICLE{Villa1976,
       author = {{Villa}, G. and {Page}, C.~G. and {Turner}, M.~J.~L. and {Cooke}, B.~A. and {Ricketts}, M.~J. and {Pounds}, K.~A. and {Adams}, D.~J.},
        title = "{The Ariel V Sky Survey Instrument and new observations of the Milky Way.}",
      journal = {\mnras},
         year = 1976,
        month = sep,
       volume = {176},
        pages = {609-620},
          doi = {10.1093/mnras/176.3.609},
       adsurl = {https://ui.adsabs.harvard.edu/abs/1976MNRAS.176..609V}
}

@ARTICLE{Brandt1990,
       author = {{Brandt}, S. and {Lund}, N. and {Rao}, A.~R.},
        title = "{The watch all-sky monitor for the granat project}",
      journal = {Advances in Space Research},
         year = 1990,
        month = jan,
       volume = {10},
       number = {2},
        pages = {239-242},
          doi = {10.1016/0273-1177(90)90148-S},
       adsurl = {https://ui.adsabs.harvard.edu/abs/1990AdSpR..10b.239B}
}

@ARTICLE{LevineRXTE1996,
       author = {{Levine}, Alan M. and {Bradt}, Hale and {Cui}, Wei and {Jernigan}, J.~G. and {Morgan}, Edward H. and {Remillard}, Ronald and {Shirey}, Robert E. and {Smith}, Donald A.},
        title = "{First Results from the All-Sky Monitor on the Rossi X-Ray Timing Explorer}",
      journal = {\apjl},
         year = 1996,
        month = sep,
       volume = {469},
        pages = {L33},
          doi = {10.1086/310260},
archivePrefix = {arXiv},
       eprint = {astro-ph/9608109},
 primaryClass = {astro-ph},
       adsurl = {https://ui.adsabs.harvard.edu/abs/1996ApJ...469L..33L}
}

@ARTICLE{Winkler2003,
       author = {{Winkler}, C. and {Courvoisier}, T.~J. -L. and {Di Cocco}, G. and {Gehrels}, N. and {Gim{\'e}nez}, A. and {Grebenev}, S. and {Hermsen}, W. and {Mas-Hesse}, J.~M. and {Lebrun}, F. and {Lund}, N. and {Palumbo}, G.~G.~C. and {Paul}, J. and {Roques}, J. -P. and {Schnopper}, H. and {Sch{\"o}nfelder}, V. and {Sunyaev}, R. and {Teegarden}, B. and {Ubertini}, P. and {Vedrenne}, G. and {Dean}, A.~J.},
        title = "{The INTEGRAL mission}",
      journal = {\aap},
         year = 2003,
        month = nov,
       volume = {411},
        pages = {L1-L6},
          doi = {10.1051/0004-6361:20031288},
       adsurl = {https://ui.adsabs.harvard.edu/abs/2003A&A...411L...1W}
}

@ARTICLE{BAT2005,
       author = {{Barthelmy}, Scott D. and {Barbier}, Louis M. and {Cummings}, Jay R. and {Fenimore}, Ed E. and {Gehrels}, Neil and {Hullinger}, Derek and {Krimm}, Hans A. and {Markwardt}, Craig B. and {Palmer}, David M. and {Parsons}, Ann and {Sato}, Goro and {Suzuki}, Masaya and {Takahashi}, Tadayuki and {Tashiro}, Makota and {Tueller}, Jack},
        title = "{The Burst Alert Telescope (BAT) on the SWIFT Midex Mission}",
      journal = {\ssr},
         year = 2005,
        month = oct,
       volume = {120},
       number = {3-4},
        pages = {143-164},
          doi = {10.1007/s11214-005-5096-3},
archivePrefix = {arXiv},
       eprint = {astro-ph/0507410},
 primaryClass = {astro-ph},
       adsurl = {https://ui.adsabs.harvard.edu/abs/2005SSRv..120..143B}
}

@ARTICLE{MAXI2009,
       author = {{Matsuoka}, Masaru and {Kawasaki}, Kazuyoshi and {Ueno}, Shiro and {Tomida}, Hiroshi and {Kohama}, Mitsuhiro and {Suzuki}, Motoko and {Adachi}, Yasuki and {Ishikawa}, Masaki and {Mihara}, Tatehiro and {Sugizaki}, Mutsumi and {Isobe}, Naoki and {Nakagawa}, Yujin and {Tsunemi}, Hiroshi and {Miyata}, Emi and {Kawai}, Nobuyuki and {Kataoka}, Jun and {Morii}, Mikio and {Yoshida}, Atsumasa and {Negoro}, Hitoshi and {Nakajima}, Motoki and {Ueda}, Yoshihiro and {Chujo}, Hirotaka and {Yamaoka}, Kazutaka and {Yamazaki}, Osamu and {Nakahira}, Satoshi and {You}, Tetsuya and {Ishiwata}, Ryoji and {Miyoshi}, Sho and {Eguchi}, Satoshi and {Hiroi}, Kazuo and {Katayama}, Haruyoshi and {Ebisawa}, Ken},
        title = "{The MAXI Mission on the ISS: Science and Instruments for Monitoring All-Sky X-Ray Images}",
      journal = {\pasj},
         year = 2009,
        month = oct,
       volume = {61},
        pages = {999},
          doi = {10.1093/pasj/61.5.999},
archivePrefix = {arXiv},
       eprint = {0906.0631},
 primaryClass = {astro-ph.IM},
       adsurl = {https://ui.adsabs.harvard.edu/abs/2009PASJ...61..999M}
}

@ARTICLE{2024ATel16765....1L,
       author = {{Liu}, H.~Y. and {Yang}, H.~N. and {Li}, D.~Y. and {Zhao}, D.~H. and {Yuan}, W. and {Sun}, H. and {Liu}, Y. and {Jin}, C.~C. and {Ling}, Z.~X. and {Zhang}, C. and {Cheng}, H.~Q. and {Chen}, W. and {Cui}, C.~Z. and {Fan}, D.~W. and {Hu}, H.~B. and {Hu}, J.~W. and {Huang}, M.~H. and {Liu}, M.~J. and {Lv}, Z.~Z. and {Lian}, T.~Y. and {Mao}, X. and {Pan}, H.~W. and {Pan}, X. and {Wang}, W.~X. and {Wang}, Y.~L. and {Wu}, Q.~Y. and {Xu}, X.~P. and {Xu}, Y.~F. and {Zhang}, M. and {Zhang}, W.~D. and {Zhang}, W.~J. and {Zhang}, Z. and {Chen}, Y. and {Jia}, S.~M. and {Cui}, W.~W. and {Feng}, H. and {Han}, D.~W. and {Li}, C.~K. and {Song}, L.~M. and {Zhao}, X.~F. and {Zhang}, J. and {Zhang}, S.~N. and {Kuulkers}, E. and {Santovincenzo}, A. and {O'Brien}, P. and {Nandra}, K. and {Rau}, A. and {Cordi}, B.},
        title = "{EP-WXT detection of an X-ray transient EP240809a and Swift follow-up observation}",
      journal = {The Astronomer's Telegram},
         year = 2024,
        month = aug,
       volume = {16765},
        pages = {1},
       adsurl = {https://ui.adsabs.harvard.edu/abs/2024ATel16765....1L}
}

@ARTICLE{2025ATel17083....1L,
       author = {{Li}, D.~Y. and {Shui}, Q.~C. and {Huang}, M.~Q. and {Liu}, Y. and {Yuan}, W.},
        title = "{Einstein Probe detection of an X-ray transient EP250315b}",
      journal = {The Astronomer's Telegram},
         year = 2025,
        month = mar,
       volume = {17083},
        pages = {1},
       adsurl = {https://ui.adsabs.harvard.edu/abs/2025ATel17083....1L}
}

@ARTICLE{Grimm_2002,
       author = {{Grimm}, H. -J. and {Gilfanov}, M. and {Sunyaev}, R.},
        title = "{The Milky Way in X-rays for an outside observer. Log(N)-Log(S) and luminosity function of X-ray binaries from RXTE/ASM data}",
      journal = {\aap},
         year = 2002,
        month = sep,
       volume = {391},
        pages = {923-944},
          doi = {10.1051/0004-6361:20020826},
archivePrefix = {arXiv},
       eprint = {astro-ph/0109239},
 primaryClass = {astro-ph},
       adsurl = {https://ui.adsabs.harvard.edu/abs/2002A&A...391..923G}
}

@ARTICLE{Chen_FRED,
       author = {{Chen}, Wan and {Shrader}, C.~R. and {Livio}, Mario},
        title = "{The Properties of X-Ray and Optical Light Curves of X-Ray Novae}",
      journal = {\apj},
         year = 1997,
        month = dec,
       volume = {491},
       number = {1},
        pages = {312-338},
          doi = {10.1086/304921},
archivePrefix = {arXiv},
       eprint = {astro-ph/9707138},
 primaryClass = {astro-ph},
       adsurl = {https://ui.adsabs.harvard.edu/abs/1997ApJ...491..312C}
}

@ARTICLE{Remillard_review,
       author = {{Remillard}, Ronald A. and {McClintock}, Jeffrey E.},
        title = "{X-Ray Properties of Black-Hole Binaries}",
      journal = {\araa},
         year = 2006,
        month = sep,
       volume = {44},
       number = {1},
        pages = {49-92},
          doi = {10.1146/annurev.astro.44.051905.092532},
archivePrefix = {arXiv},
       eprint = {astro-ph/0606352},
 primaryClass = {astro-ph},
       adsurl = {https://ui.adsabs.harvard.edu/abs/2006ARA&A..44...49R}
}

@ARTICLE{Wijnands_2006,
       author = {{Wijnands}, R. and {in't Zand}, J.~J.~M. and {Rupen}, M. and {Maccarone}, T. and {Homan}, J. and {Cornelisse}, R. and {Fender}, R. and {Grindlay}, J. and {van der Klis}, M. and {Kuulkers}, E. and {Markwardt}, C.~B. and {Miller-Jones}, J.~C.~A. and {Wang}, Q.~D.},
        title = "{The XMM-Newton/Chandra monitoring campaign of the Galactic center region. Description of the program and preliminary results}",
      journal = {\aap},
         year = 2006,
        month = apr,
       volume = {449},
       number = {3},
        pages = {1117-1127},
          doi = {10.1051/0004-6361:20054129},
archivePrefix = {arXiv},
       eprint = {astro-ph/0508648},
 primaryClass = {astro-ph},
       adsurl = {https://ui.adsabs.harvard.edu/abs/2006A&A...449.1117W}
}

@ARTICLE{Stoop2021,
       author = {{Stoop}, M. and {van den Eijnden}, J. and {Degenaar}, N. and {Bahramian}, A. and {Swihart}, S.~J. and {Strader}, J. and {Jim{\'e}nez-Ibarra}, F. and {Mu{\~n}oz-Darias}, T. and {Armas Padilla}, M. and {Shaw}, A.~W. and {Maccarone}, T.~J. and {Wijnands}, R. and {Russell}, T.~D. and {Hern{\'a}ndez Santisteban}, J.~V. and {Miller-Jones}, J.~C.~A. and {Russell}, D.~M. and {Maitra}, D. and {Heinke}, C.~O. and {Sivakoff}, G.~R. and {Lewis}, F. and {Bramich}, D.~M.},
        title = "{Multiwavelength observations reveal a faint candidate black hole X-ray binary in IGR J17285-2922}",
      journal = {\mnras},
         year = 2021,
        month = oct,
       volume = {507},
       number = {1},
        pages = {330-349},
          doi = {10.1093/mnras/stab2127},
       adsurl = {https://ui.adsabs.harvard.edu/abs/2021MNRAS.507..330S}
}

@ARTICLE{2018IGRJ17062-6143 ,
       author = {{van den Eijnden}, J. and {Degenaar}, N. and {Pinto}, C. and {Patruno}, A. and {Wette}, K. and {Messenger}, C. and {Hern{\'a}ndez Santisteban}, J.~V. and {Wijnands}, R. and {Miller}, J.~M. and {Altamirano}, D. and {Paerels}, F. and {Chakrabarty}, D. and {Fabian}, A.~C.},
        title = "{The very faint X-ray binary IGR J17062-6143: a truncated disc, no pulsations, and a possible outflow}",
      journal = {\mnras},
         year = 2018,
        month = apr,
       volume = {475},
       number = {2},
        pages = {2027-2044},
          doi = {10.1093/mnras/stx3224},
       adsurl = {https://ui.adsabs.harvard.edu/abs/2018MNRAS.475.2027V}
}

@ARTICLE{2014MNRAS.442..372K,
       author = {{Koch}, E.~W. and {Bahramian}, A. and {Heinke}, C.~O. and {Mori}, K. and {Rea}, N. and {Degenaar}, N. and {Haggard}, D. and {Wijnands}, R. and {Ponti}, G. and {Miller}, J.~M. and {Yusef-Zadeh}, F. and {Dufour}, F. and {Cotton}, W.~D. and {Baganoff}, F.~K. and {Reynolds}, M.~T.},
        title = "{The 2013 outburst of a transient very faint X-ray binary, 23 arcsec from Sgr A*}",
      journal = {\mnras},
         year = 2014,
        month = jul,
       volume = {442},
       number = {1},
        pages = {372-381},
          doi = {10.1093/mnras/stu887},
archivePrefix = {arXiv},
       eprint = {1405.0267},
 primaryClass = {astro-ph.HE},
       adsurl = {https://ui.adsabs.harvard.edu/abs/2014MNRAS.442..372K}
}

@ARTICLE{2022svom,
       author = {{Atteia}, J.-L. and {Cordier}, B. and {Wei}, J.},
        title = "{The SVOM mission}",
      journal = {International Journal of Modern Physics D},
         year = 2022,
        month = apr,
       volume = {31},
       number = {5},
          eid = {2230008},
        pages = {2230008},
          doi = {10.1142/S0218271822300087},
archivePrefix = {arXiv},
       eprint = {2203.10962},
 primaryClass = {astro-ph.IM},
       adsurl = {https://ui.adsabs.harvard.edu/abs/2022IJMPD..3130008A}
}

@ARTICLE{Cash,
       author = {{Cash}, W.},
        title = "{Parameter estimation in astronomy through application of the likelihood ratio.}",
      journal = {\apj},
         year = 1979,
        month = mar,
       volume = {228},
        pages = {939-947},
          doi = {10.1086/156922},
       adsurl = {https://ui.adsabs.harvard.edu/abs/1979ApJ...228..939C}
}

@ARTICLE{Evans2009,
       author = {{Evans}, P.~A. and {Beardmore}, A.~P. and {Page}, K.~L. and {Osborne}, J.~P. and {O'Brien}, P.~T. and {Willingale}, R. and {Starling}, R.~L.~C. and {Burrows}, D.~N. and {Godet}, O. and {Vetere}, L. and {Racusin}, J. and {Goad}, M.~R. and {Wiersema}, K. and {Angelini}, L. and {Capalbi}, M. and {Chincarini}, G. and {Gehrels}, N. and {Kennea}, J.~A. and {Margutti}, R. and {Morris}, D.~C. and {Mountford}, C.~J. and {Pagani}, C. and {Perri}, M. and {Romano}, P. and {Tanvir}, N.},
        title = "{Methods and results of an automatic analysis of a complete sample of Swift-XRT observations of GRBs}",
      journal = {\mnras},
         year = 2009,
        month = aug,
       volume = {397},
       number = {3},
        pages = {1177-1201},
          doi = {10.1111/j.1365-2966.2009.14913.x},
archivePrefix = {arXiv},
       eprint = {0812.3662},
 primaryClass = {astro-ph},
       adsurl = {https://ui.adsabs.harvard.edu/abs/2009MNRAS.397.1177E}
}

@INPROCEEDINGS{arnaud96,
       author = {{Arnaud}, K.~A.},
        title = "{XSPEC: The First Ten Years}",
    booktitle = {Astronomical Data Analysis Software and Systems V},
         year = 1996,
       editor = {{Jacoby}, George H. and {Barnes}, Jeannette},
       series = {Astronomical Society of the Pacific Conference Series},
       volume = {101},
        month = jan,
        pages = {17},
       adsurl = {https://ui.adsabs.harvard.edu/abs/1996ASPC..101...17A}
}

@ARTICLE{2025ApJS..280...15W,
       author = {{Wang}, Tao and {Yuan}, Haibo and {Chen}, Bingqiu and {Xiang}, Maosheng and {Zhang}, Ruoyi and {Huang}, Bowen and {Gu}, Hongrui and {Wang}, Shuaicong and {Li}, Jiawei},
        title = "{An All-sky 3D Dust Map Based on Gaia and LAMOST}",
      journal = {\apjs},
         year = 2025,
        month = sep,
       volume = {280},
       number = {1},
          eid = {15},
        pages = {15},
          doi = {10.3847/1538-4365/adea39},
archivePrefix = {arXiv},
       eprint = {2509.07640},
 primaryClass = {astro-ph.GA},
       adsurl = {https://ui.adsabs.harvard.edu/abs/2025ApJS..280...15W}
}

@ARTICLE{2006MNRAS.371.1334R,
       author = {{Russell}, D.~M. and {Fender}, R.~P. and {Hynes}, R.~I. and {Brocksopp}, C. and {Homan}, J. and {Jonker}, P.~G. and {Buxton}, M.~M.},
        title = "{Global optical/infrared-X-ray correlations in X-ray binaries: quantifying disc and jet contributions}",
      journal = {\mnras},
         year = 2006,
        month = sep,
       volume = {371},
       number = {3},
        pages = {1334-1350},
          doi = {10.1111/j.1365-2966.2006.10756.x},
archivePrefix = {arXiv},
       eprint = {astro-ph/0606721},
 primaryClass = {astro-ph},
       adsurl = {https://ui.adsabs.harvard.edu/abs/2006MNRAS.371.1334R}
}

@PROCEEDINGS{1995xrbi.nasa.....L,
        title = "{X-ray binaries}",
    booktitle = {X-ray Binaries},
         year = 1995,
       editor = {{Lewin}, Walter H.~G. and {van Paradijs}, Jan and {van den Heuvel}, Edward P.~J.},
        month = jan,
       adsurl = {https://ui.adsabs.harvard.edu/abs/1995xrbi.nasa.....L}
}

@ARTICLE{1989ApJ...345..245C,
       author = {{Cardelli}, Jason A. and {Clayton}, Geoffrey C. and {Mathis}, John S.},
        title = "{The Relationship between Infrared, Optical, and Ultraviolet Extinction}",
      journal = {\apj},
         year = 1989,
        month = oct,
       volume = {345},
        pages = {245},
          doi = {10.1086/167900},
       adsurl = {https://ui.adsabs.harvard.edu/abs/1989ApJ...345..245C}
}

@ARTICLE{2023A&A...674A...1G,
       author = {{Gaia Collaboration} and {Vallenari}, A. and {Brown}, A.~G.~A. and {Prusti}, T. and {de Bruijne}, J.~H.~J. and {Arenou}, F. and {Babusiaux}, C. and {Biermann}, M. and {Creevey}, O.~L. and {Ducourant}, C. and {Evans}, D.~W. and {Eyer}, L. and {Guerra}, R. and {Hutton}, A. and {Jordi}, C. and {Klioner}, S.~A. and {Lammers}, U.~L. and {Lindegren}, L. and {Luri}, X. and {Mignard}, F. and {Panem}, C. and {Pourbaix}, D. and {Randich}, S. and {Sartoretti}, P. and {Soubiran}, C. and {Tanga}, P. and {Walton}, N.~A. and {Bailer-Jones}, C.~A.~L. and {Bastian}, U. and {Drimmel}, R. and {Jansen}, F. and {Katz}, D. and {Lattanzi}, M.~G. and {van Leeuwen}, F. and {Bakker}, J. and {Cacciari}, C. and {Casta{\~n}eda}, J. and {De Angeli}, F. and {Fabricius}, C. and {Fouesneau}, M. and {Fr{\'e}mat}, Y. and {Galluccio}, L. and {Guerrier}, A. and {Heiter}, U. and {Masana}, E. and {Messineo}, R. and {Mowlavi}, N. and {Nicolas}, C. and {Nienartowicz}, K. and {Pailler}, F. and {Panuzzo}, P. and {Riclet}, F. and {Roux}, W. and {Seabroke}, G.~M. and {Sordo}, R. and {Th{\'e}venin}, F. and {Gracia-Abril}, G. and {Portell}, J. and {Teyssier}, D. and {Altmann}, M. and {Andrae}, R. and {Audard}, M. and {Bellas-Velidis}, I. and {Benson}, K. and {Berthier}, J. and {Blomme}, R. and {Burgess}, P.~W. and {Busonero}, D. and {Busso}, G. and {C{\'a}novas}, H. and {Carry}, B. and {Cellino}, A. and {Cheek}, N. and {Clementini}, G. and {Damerdji}, Y. and {Davidson}, M. and {de Teodoro}, P. and {Nu{\~n}ez Campos}, M. and {Delchambre}, L. and {Dell'Oro}, A. and {Esquej}, P. and {Fern{\'a}ndez-Hern{\'a}ndez}, J. and {Fraile}, E. and {Garabato}, D. and {Garc{\'\i}a-Lario}, P. and {Gosset}, E. and {Haigron}, R. and {Halbwachs}, J.-L. and {Hambly}, N.~C. and {Harrison}, D.~L. and {Hern{\'a}ndez}, J. and {Hestroffer}, D. and {Hodgkin}, S.~T. and {Holl}, B. and {Jan{\ss}en}, K. and {Jevardat de Fombelle}, G. and {Jordan}, S. and {Krone-Martins}, A. and {Lanzafame}, A.~C. and {L{\"o}ffler}, W. and {Marchal}, O. and {Marrese}, P.~M. and {Moitinho}, A. and {Muinonen}, K. and {Osborne}, P. and {Pancino}, E. and {Pauwels}, T. and {Recio-Blanco}, A. and {Reyl{\'e}}, C. and {Riello}, M. and {Rimoldini}, L. and {Roegiers}, T. and {Rybizki}, J. and {Sarro}, L.~M. and {Siopis}, C. and {Smith}, M. and {Sozzetti}, A. and {Utrilla}, E. and {van Leeuwen}, M. and {Abbas}, U. and {{\'A}brah{\'a}m}, P. and {Abreu Aramburu}, A. and {Aerts}, C. and {Aguado}, J.~J. and {Ajaj}, M. and {Aldea-Montero}, F. and {Altavilla}, G. and {{\'A}lvarez}, M.~A. and {Alves}, J. and {Anders}, F. and {Anderson}, R.~I. and {Anglada Varela}, E. and {Antoja}, T. and {Baines}, D. and {Baker}, S.~G. and {Balaguer-N{\'u}{\~n}ez}, L. and {Balbinot}, E. and {Balog}, Z. and {Barache}, C. and {Barbato}, D. and {Barros}, M. and {Barstow}, M.~A. and {Bartolom{\'e}}, S. and {Bassilana}, J.-L. and {Bauchet}, N. and {Becciani}, U. and {Bellazzini}, M. and {Berihuete}, A. and {Bernet}, M. and {Bertone}, S. and {Bianchi}, L. and {Binnenfeld}, A. and {Blanco-Cuaresma}, S. and {Blazere}, A. and {Boch}, T. and {Bombrun}, A. and {Bossini}, D. and {Bouquillon}, S. and {Bragaglia}, A. and {Bramante}, L. and {Breedt}, E. and {Bressan}, A. and {Brouillet}, N. and {Brugaletta}, E. and {Bucciarelli}, B. and {Burlacu}, A. and {Butkevich}, A.~G. and {Buzzi}, R. and {Caffau}, E. and {Cancelliere}, R. and {Cantat-Gaudin}, T. and {Carballo}, R. and {Carlucci}, T. and {Carnerero}, M.~I. and {Carrasco}, J.~M. and {Casamiquela}, L. and {Castellani}, M. and {Castro-Ginard}, A. and {Chaoul}, L. and {Charlot}, P. and {Chemin}, L. and {Chiaramida}, V. and {Chiavassa}, A. and {Chornay}, N. and {Comoretto}, G. and {Contursi}, G. and {Cooper}, W.~J. and {Cornez}, T. and {Cowell}, S. and {Crifo}, F. and {Cropper}, M. and {Crosta}, M. and {Crowley}, C. and {Dafonte}, C. and {Dapergolas}, A. and {David}, M. and {David}, P. and {de Laverny}, P. and {De Luise}, F. and {De March}, R.},
        title = "{Gaia Data Release 3. Summary of the content and survey properties}",
      journal = {\aap},
         year = 2023,
        month = jun,
       volume = {674},
          eid = {A1},
        pages = {A1},
          doi = {10.1051/0004-6361/202243940},
archivePrefix = {arXiv},
       eprint = {2208.00211},
 primaryClass = {astro-ph.GA},
       adsurl = {https://ui.adsabs.harvard.edu/abs/2023A&A...674A...1G}
}

@ARTICLE{2013MNRAS.430.1694D,
       author = {{Dauser}, T. and {Garcia}, J. and {Wilms}, J. and {B{\"o}ck}, M. and {Brenneman}, L.~W. and {Falanga}, M. and {Fukumura}, K. and {Reynolds}, C.~S.},
        title = "{Irradiation of an accretion disc by a jet: general properties and implications for spin measurements of black holes}",
      journal = {\mnras},
         year = 2013,
        month = apr,
       volume = {430},
       number = {3},
        pages = {1694-1708},
          doi = {10.1093/mnras/sts710},
archivePrefix = {arXiv},
       eprint = {1301.4922},
 primaryClass = {astro-ph.HE},
       adsurl = {https://ui.adsabs.harvard.edu/abs/2013MNRAS.430.1694D}
}

@ARTICLE{2014ApJ...782...76G,
       author = {{Garc{\'\i}a}, J. and {Dauser}, T. and {Lohfink}, A. and {Kallman}, T.~R. and {Steiner}, J.~F. and {McClintock}, J.~E. and {Brenneman}, L. and {Wilms}, J. and {Eikmann}, W. and {Reynolds}, C.~S. and {Tombesi}, F.},
        title = "{Improved Reflection Models of Black Hole Accretion Disks: Treating the Angular Distribution of X-Rays}",
      journal = {\apj},
         year = 2014,
        month = feb,
       volume = {782},
       number = {2},
          eid = {76},
        pages = {76},
          doi = {10.1088/0004-637X/782/2/76},
archivePrefix = {arXiv},
       eprint = {1312.3231},
 primaryClass = {astro-ph.HE},
       adsurl = {https://ui.adsabs.harvard.edu/abs/2014ApJ...782...76G}
}

@ARTICLE{2025ATel17397....1I,
       author = {{Illiano}, Giulia and {Zanon}, Arianna Miraval and {Sbarufatti}, Boris and {Papitto}, Alessandro and {Baglio}, Maria Cristina and {Ballocco}, Caterina},
        title = "{Swift follow-up of the X-ray transient EP250916a}",
      journal = {The Astronomer's Telegram},
         year = 2025,
        month = sep,
       volume = {17397},
        pages = {1},
       adsurl = {https://ui.adsabs.harvard.edu/abs/2025ATel17397....1I}
}

@software{Ransom2011,
       author = {{Ransom}, Scott},
        title = "{PRESTO: PulsaR Exploration and Search TOolkit}",
 howpublished = {Astrophysics Source Code Library, record ascl:1107.017},
         year = 2011,
        month = jul,
          eid = {ascl:1107.017},
archivePrefix = {ascl},
       eprint = {1107.017},
       adsurl = {https://ui.adsabs.harvard.edu/abs/2011ascl.soft07017R}
}

@ARTICLE{2017MNRAS.466..194B,
       author = {{Burke}, M.~J. and {Gilfanov}, M. and {Sunyaev}, R.},
        title = "{A dichotomy between the hard state spectral properties of black hole and neutron star X-ray binaries}",
      journal = {\mnras},
         year = 2017,
        month = apr,
       volume = {466},
       number = {1},
        pages = {194-212},
          doi = {10.1093/mnras/stw2514},
archivePrefix = {arXiv},
       eprint = {1609.09511},
 primaryClass = {astro-ph.HE},
       adsurl = {https://ui.adsabs.harvard.edu/abs/2017MNRAS.466..194B}
}

@ARTICLE{2025ApJS..279...57H,
       author = {{Heinke}, Craig O. and {Zheng}, Junwen and {Maccarone}, Thomas J. and {Degenaar}, Nathalie and {Bahramian}, Arash and {Sivakoff}, Gregory R. and {Toor}, Simrat},
        title = "{Catalog of Outbursts of Neutron Star Low-mass X-Ray Binaries}",
      journal = {\apjs},
         year = 2025,
        month = aug,
       volume = {279},
       number = {2},
          eid = {57},
        pages = {57},
          doi = {10.3847/1538-4365/ade99a},
archivePrefix = {arXiv},
       eprint = {2407.18867},
 primaryClass = {astro-ph.HE},
       adsurl = {https://ui.adsabs.harvard.edu/abs/2025ApJS..279...57H}
}

@ARTICLE{2022MNRAS.515.3838M,
       author = {{Marino}, A. and {Anitra}, A. and {Mazzola}, S.~M. and {Di Salvo}, T. and {Sanna}, A. and {Bult}, P. and {Guillot}, S. and {Mancuso}, G. and {Ng}, M. and {Riggio}, A. and {Albayati}, A.~C. and {Altamirano}, D. and {Arzoumanian}, Z. and {Burderi}, L. and {Cabras}, C. and {Chakrabarty}, D. and {Deiosso}, N. and {Gendreau}, K.~C. and {Iaria}, R. and {Manca}, A. and {Strohmayer}, T.~E.},
        title = "{Outflows and spectral evolution in the eclipsing AMXP SWIFT J1749.4-2807 with NICER, XMM-Newton, and NuSTAR}",
      journal = {\mnras},
         year = 2022,
        month = sep,
       volume = {515},
       number = {3},
        pages = {3838-3852},
          doi = {10.1093/mnras/stac2038},
archivePrefix = {arXiv},
       eprint = {2207.08637},
 primaryClass = {astro-ph.HE},
       adsurl = {https://ui.adsabs.harvard.edu/abs/2022MNRAS.515.3838M}
}

@ARTICLE{2022ApJ...927..190P,
       author = {{Pike}, Sean N. and {Negoro}, Hitoshi and {Tomsick}, John A. and {Bachetti}, Matteo and {Brumback}, McKinley and {Connors}, Riley M.~T. and {Garc{\'\i}a}, Javier A. and {Grefenstette}, Brian and {Hare}, Jeremy and {Harrison}, Fiona A. and {Jaodand}, Amruta and {Ludlam}, R.~M. and {Mastroserio}, Guglielmo and {Mihara}, Tatehiro and {Shidatsu}, Megumi and {Sugizaki}, Mutsumi and {Takagi}, Ryohei},
        title = "{MAXI and NuSTAR Observations of the Faint X-Ray Transient MAXI J1848-015 in the GLIMPSE-C01 Cluster}",
      journal = {\apj},
         year = 2022,
        month = mar,
       volume = {927},
       number = {2},
          eid = {190},
        pages = {190},
          doi = {10.3847/1538-4357/ac5258},
archivePrefix = {arXiv},
       eprint = {2202.02883},
 primaryClass = {astro-ph.HE},
       adsurl = {https://ui.adsabs.harvard.edu/abs/2022ApJ...927..190P}
}

@INCOLLECTION{2023hxga.book..139B,
       author = {{Bu}, Qingcui and {Zhang}, Shuangnan},
        title = "{Black Holes: Accretion Processes in X-ray Binaries}",
    booktitle = {Handbook of X-ray and Gamma-ray Astrophysics},
         year = 2023,
          eid = {139},
        pages = {139},
          doi = {10.1007/978-981-16-4544-0_99-1},
       adsurl = {https://ui.adsabs.harvard.edu/abs/2023hxga.book..139B}
}

@ARTICLE{2007A&ARv..15....1D,
       author = {{Done}, Chris and {Gierli{\'n}ski}, Marek and {Kubota}, Aya},
        title = "{Modelling the behaviour of accretion flows in X-ray binaries. Everything you always wanted to know about accretion but were afraid to ask}",
      journal = {\aapr},
         year = 2007,
        month = dec,
       volume = {15},
       number = {1},
        pages = {1-66},
          doi = {10.1007/s00159-007-0006-1},
archivePrefix = {arXiv},
       eprint = {0708.0148},
 primaryClass = {astro-ph},
       adsurl = {https://ui.adsabs.harvard.edu/abs/2007A&ARv..15....1D}
}

\begin{appendix}

\section{Data reduction}
\label{data_reduction}

\subsection{EP}
To provide full coverage of the outburst, in addition to the follow-up observations (Tab.~\ref{tab:obs_info}), we also use survey data from EP/WXT.  The WXT spectra were extracted using the standard wxtpipeline pipeline (Liu et al., in prep.) and the latest calibration database (CALDB), which is based on results from ground calibration experiments \citep{2025ExA....60...15C}. Due to the limited photon statistics of the WXT data, the spectra were grouped to a minimum of one count per bin using grppha in HEASoft v6.34, and spectral fitting was performed using the Cash statistic (C--stat; \citealt{Cash}).

Owing to solar angle constraints, EP/FXT observations were carried out on 2025 September 17 (upper-right panel of Fig.~\ref{src_image}) and on 2026 April 22. For the first observation, we analyzed only the FXT module B (FXT-B) data taken in Partial Window (PW) mode with the fxtchain tool, as the module A (FXT-A) operating in Full Frame (FF) mode was affected by severe pile-up. Both PW and FF mode data were used for the second observation. Source and background spectra were extracted from a circular region with a radius of 90\arcsec\ and from an annulus spanning 150\arcsec--240\arcsec\ centered on the source position, respectively. The spectra were grouped to a minimum of 25 counts per bin using grppha.

\subsection{NuSTAR}
NuSTAR observed this source on 2025 September 24 (ObsID 91101336002; $\sim21$~ks). Data were processed with nupipeline and CALDB version 20250922. Source events were extracted from a circular region of 60\arcsec\ radius centered on the source, and background events were extracted from a rectangular region free of stray light and sufficiently distant from the source. The spectra, light curves, and response files were produced using nuproducts, and the spectra were grouped to a minimum of 25 counts per bin.

\subsection{Swift}

Swift/XRT performed high-cadence monitoring of EP~J175257.3--351923 starting on 2025 September 17 and continuing for approximately one month, with a total of 21 observations. The spectra were extracted using the online tool provided by \citet{Evans2009}\footnote{\url{https://www.swift.ac.uk/user_objects/}}.
For spectra with more than 1500 total counts, the data were binned to $\geq 25$ counts per bin and fitted using $\chi^2$ statistics. For spectra with fewer counts, the data were grouped to a minimum of one count per bin and fitted using Cash statistic.

The Ultra-Violet/Optical Telescope (UVOT) monitored the source contemporaneously with XRT. Some observations were performed only in the $V$-band, while others included both the $B$- and $V$-bands. We analyzed the Level 3 data products.

For potential optical counterparts (see Sect.~\ref{Opt}), aperture photometry was performed using uvotsource with a 1\arcsec\ aperture and a nearby background region of 5\arcsec\ radius free of sources. While this small aperture is suboptimal for absolute photometry, it was chosen to minimize contamination from the crowded stellar field. As our analysis focuses on the evolution of optical flux during the outburst for counterpart identification, uncertainties in the aperture correction do not affect the observed trends. To estimate the upper limit of the optical emission from EP~J175257.3--351923, an aperture of 2.2\arcsec\ (corresponding to the Swift/XRT localization uncertainty) and a circular background of 5\arcsec\ radius were used. We note that with an aperture of 2.2\arcsec, emission from nearby sources is included, yielding an overestimated upper limit, though this does not affect the constraints on the companion type based on the X-ray-to-optical flux ratio (Sec.~\ref{Conclusions}). Aperture corrections were applied using the apercorr=CURVEOFGROWTH option. All magnitudes reported in this work are given in the Vegamag system.

Aperture photometry of the three Gaia sources within or near the Swift/XRT error circle using Swift/UVOT data gives averaged magnitudes of $V = 18.19$, $18.46$, and $18.16$~mag over the outburst, broadly consistent with the $V$-band magnitudes ($V = 17.70$, $18.91$, and $18.10$~mag) derived from Gaia DR3 using the relation of \cite{2021A&A...649A...3R}. However, we need to point out that the small aperture used may introduce systematic uncertainties in the magnitudes.

\subsection{SVOM}

SVOM conducted five follow-up observations between 2025 September 17 and 21, before the source became inaccessible owing to solar constraints. The VT data from September 17 were lost due to telemetry failures.

The Microchannel X-ray Telescope (MXT; \citealt{Gotz2026}) onboard SVOM observed the source during this period. The event-mode data were processed with version 1.13 of the
MXT pipeline \citep{Maggi2026}. Source detection was performed via point-spread
function (PSF) fitting, and the pipeline was used to extract the source and background
spectra, along with the associated response files, from the event lists. Owing to the relatively high background of the MXT data, the spectra were grouped to a minimum of one count per bin using grppha and analyzed with the C--stat. 

The Visible Telescope (VT) on board SVOM also observed the source region contemporaneously with WXT, with an exposure time of 70~s per frame. Level~2 data products, including dark and flat-field corrections, were used. Aperture photometry was performed using photutils in Python, with aperture corrections applied. The source aperture sizes and background regions adopted for nearby Gaia sources and for estimating the upper limit of the optical emission from EP~J175257.3--351923 are identical to those used for Swift/UVOT. The magnitudes were first converted to the Johnson $B$-band AB system following \citet{10.1088/1674-4527/ae75d6}, and subsequently transformed to the Vegamag system using the relation of \cite{2007AJ....133..734B}.

The ECLAIRs data were also examined, but the source was not detected.

\subsection{GROND}
\label{appendix_ground}

\begin{figure}[htbp]
    \centering
    \includegraphics[width=0.9\linewidth]{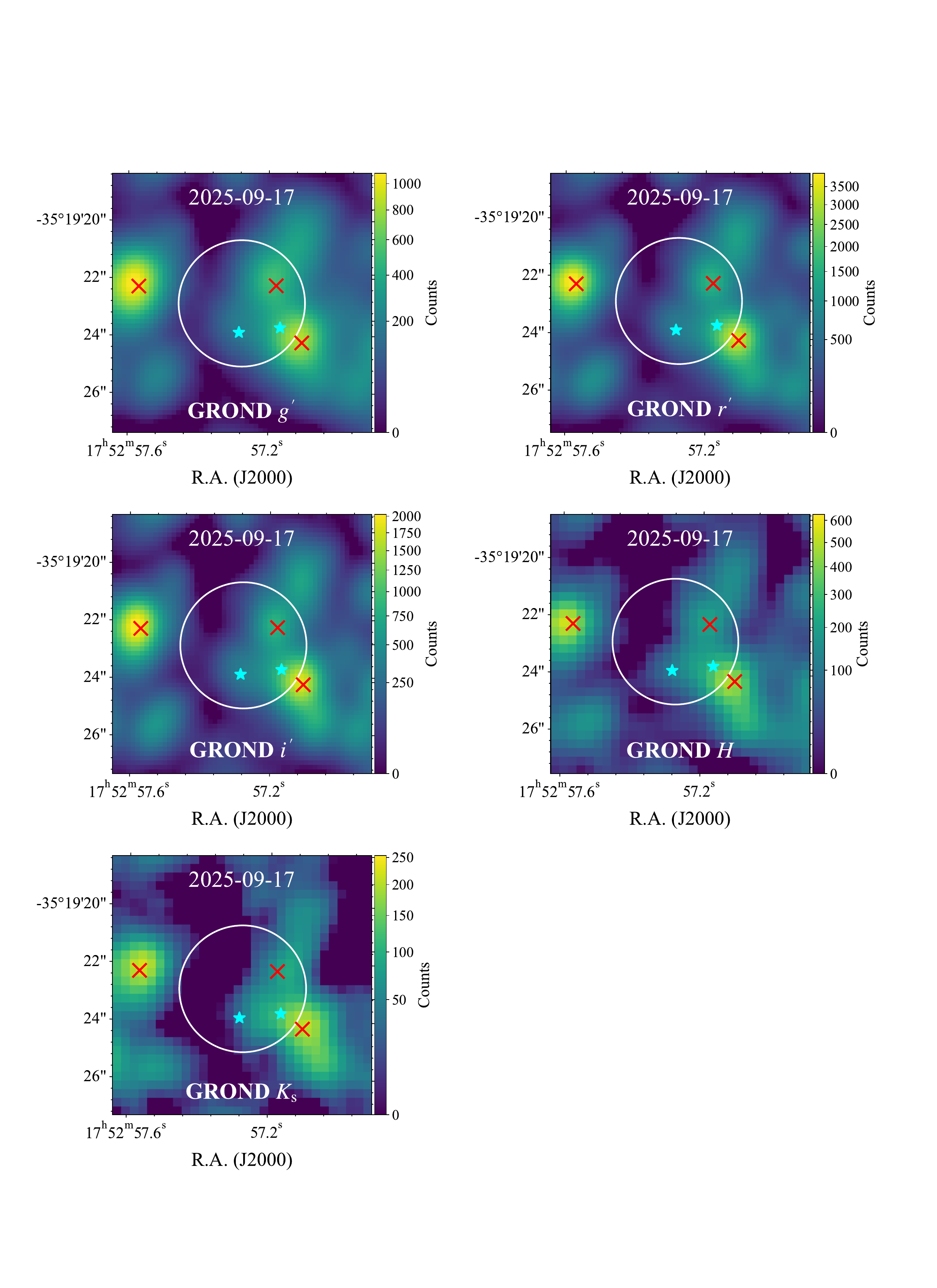}  
    \caption{GROND observations on 2025 September 17. The images have been smoothed for display purposes. Symbols follow the same convention as in Fig.~\ref{src_image}.}
    \label{image_grond}
\end{figure}

This source was observed with the GROND  mounted on the MPG 2.2\,m telescope at ESO's La Silla Observatory. Observations were carried out simultaneously in the $g^{\prime}$, $r^{\prime}$, and $i^{\prime}$ bands, as well as in the $H$ and $K_{\rm s}$ bands, on 2025 September 17 at 01:02\,UT, September 18 at 23:55\,UT, and September 27 at 01:12\,UT. The total exposure times were 33\,min in the optical and 30\,min in the near-infrared. 
The data were reduced using the standard IRAF-based GROND pipeline \citep{2008ApJ...685..376K}.

\subsection{VLASS}
\label{VLASS}
The source position was covered by VLASS on 2026 February 1. No radio counterpart is detected within the XRT error circle in the VLASS quick-look image. The local rms noise was measured within a circular region of radius $20''$ centered on the XRT position, yielding a $3\sigma$ flux density upper limit of $4.44\times10^{-4}\,\mathrm{Jy\,beam^{-1}}$ at 3\,GHz.

\section{Details of X-ray spectral fitting
\label{spec_details}}

\subsection{Swift/XRT and EP/FXT}
Between 2025 September 17 and October 8, the Swift/XRT WT-mode spectra require an absorbed power-law plus diskbb model (simftest proves the diskbb addition is statistically significant, $p < 0.01$). For the Swift/XRT PC-mode spectra, the diskbb parameters could not be constrained because of limited photon statistics after excluding central pixels to mitigate pile-up. They were therefore fixed to the values derived from temporally adjacent WT-mode observations. After October 8, a single absorbed power-law is sufficient. For the two EP/FXT observations, the spectra are best fitted with an absorbed power-law plus diskbb model.

\subsection{SVOM/MXT}
The SVOM/MXT spectra alone do not require an additional thermal component and are adequately described by a single absorbed power-law model. For consistency with the Swift/XRT and EP/FXT spectral analysis, we nevertheless adopted the same absorbed power-law plus diskbb model when fitting the MXT spectra. However, when the diskbb parameters were fixed to the values derived from adjacent Swift/XRT observations, the inferred power-law photon index became significantly softer. This behavior indicates a degeneracy between the thermal and non-thermal components within the MXT bandpass. We caution that the MXT-derived photon index should be interpreted cautiously due to these degeneracies.

\subsection{EP/WXT}
For the EP/WXT spectra during the outburst, the narrow energy coverage and limited photon statistics allowed only an absorbed power-law fit, with $N_{\rm H}$ fixed to the value derived from the adjacent EP/FXT observations and $\Gamma$ fixed at 2.

To derive the pre-outburst flux upper limit, we combined the three WXT observations covering this position obtained on 2025 September 14 and 15, yielding a total exposure time of $\sim6400$~s. The source region was defined as a circle centered on the Swift/XRT position with a radius of $9.2\,\arcmin$, while the background was extracted from a nearby source-free region with a radius of $30\,\arcmin$. The 90\% upper limit on the count rate was converted to an unabsorbed flux upper limit using the count-rate-to-flux conversion factor, assuming an absorbed power-law model with $N_{\rm H}$ derived from the first EP/FXT observation and $\Gamma=2$.

\subsection{Joint NuSTAR and Swift spectral fitting}
\begin{figure}[htbp]
    \centering
    \includegraphics[width=0.9\linewidth]{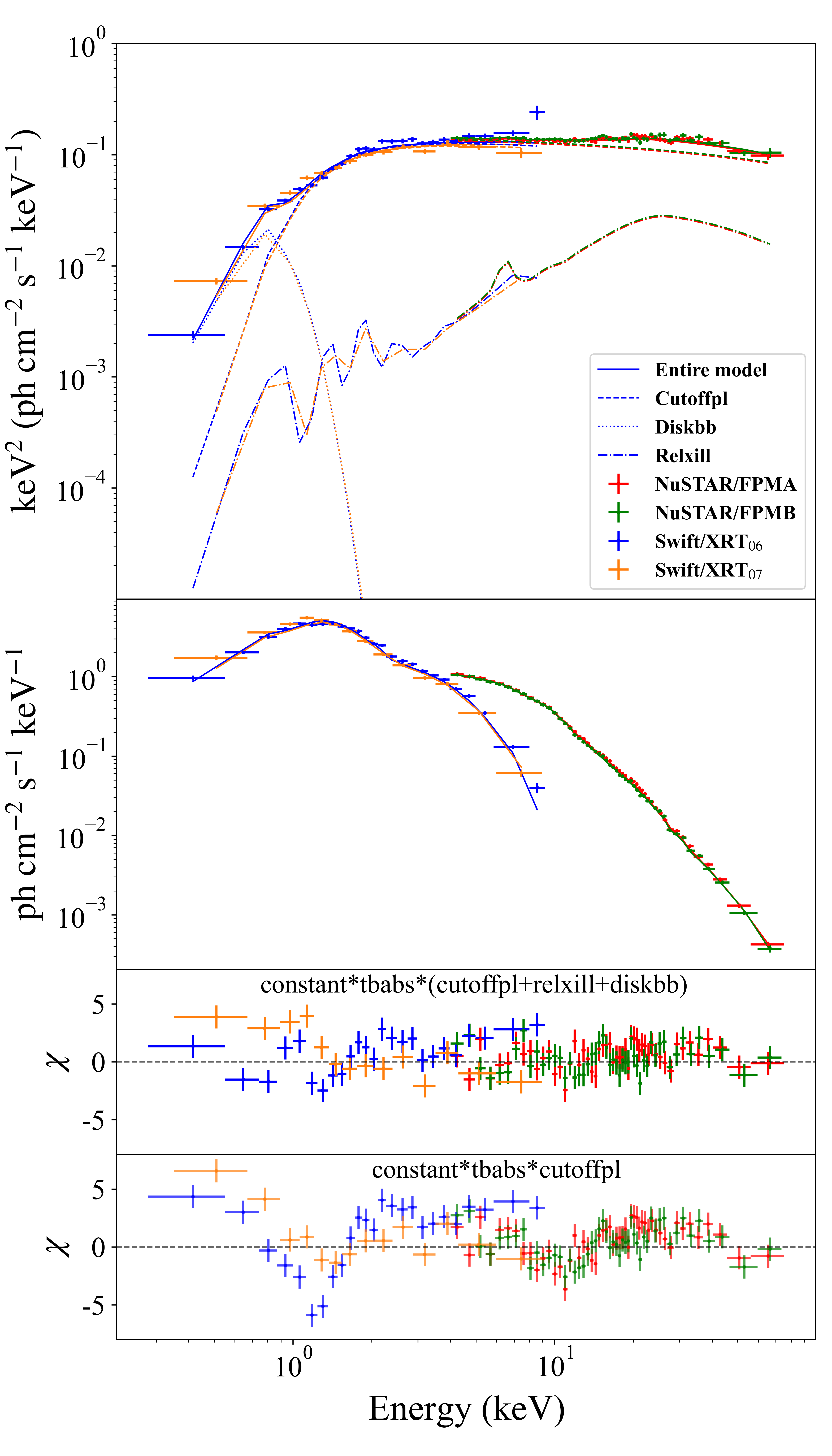}  
    \caption{
    Joint spectral fitting of the Swift/XRT (2025 September 23 and 26) and NuSTAR (2025 September 24) observations. 
Top: Unfolded spectrum in units of keV$^2$ (ph cm$^{-2}$ s$^{-1}$ keV$^{-1}$) with the best-fit model tbabs*(cutoffpl+relxill+diskbb). 
Second: Unfolded spectrum in units of ph cm$^{-2}$ s$^{-1}$ keV$^{-1}$. 
Third and bottom: Residuals for tbabs*(cutoffpl+relxill+diskbb) and tbabs*cutoffpl, respectively. The spectra are rebinned for display purposes.
    }
    \label{fig_spec}
\end{figure}

The NuSTAR spectra obtained on 2025 September 24 were fitted jointly with two quasi-simultaneous Swift/XRT observations taken on September 23 and 26 to improve soft-band photon statistics. 
An absorbed cutoff power-law improves the fit over a simple power-law, reducing $\chi^2/\mathrm{d.o.f.}$ from $2158/1669$ to $2093/1668$, yielding a cutoff energy of $154^{+41}_{-27}$~keV.
However, the cutoff power-law model still leaves soft X-ray residuals and a weak broad hump at 20--40\,keV (Fig.~\ref{fig_spec}), suggesting the presence of additional disk and reflection components. Consequently, we used constant*tbabs*(cutoffpl+relxill+diskbb) to fit the spectra, where relxill is a model for relativistic reflection on an accretion disk \citep{2013MNRAS.430.1694D,2014ApJ...782...76G}. As several parameters of relxill could not be well constrained, we fixed the black hole spin ($a$) at 0.998, the inner disk radius ($R_{\rm in}$) at $100\,R_{\rm g}$ (as suggested by the long-term evolution of the diskbb parameters; Fig.~\ref{fig:long_lc}), and the emissivity indices ($q_1$ and $q_2$) at 3, with the reflection fraction fixed at $-1$ so that relxill accounts only for the reflected component. This model provided a good fit with $\chi^2/\rm{d.o.f.}=1853/1662$, with the best-fit parameters listed in Tab.~\ref{tab_fit_spec}, yielding a cut-off energy of $217_{-50}^{+72}$~keV and an inclination angle of $58^{+16}_{-31}\,^\circ$.

\section{Long-term light-curve modeling and binary constraints}

\subsection{FRED modeling}
\label{fred}

We fit the multi-instrument long-term X-ray light curve, excluding the pre-outburst upper limits, using a modified FRED model \citep{2007MNRAS.374..466P} parameterized as
\begin{equation}
F(t)=
\begin{cases}
A \exp[(t-t_{\rm p})/\tau_{\rm r}], & t < t_{\rm p}, \\
(A-F_d)\exp[-(t-t_{\rm p})/\tau_{\rm d}] + F_d, & t \ge t_{\rm p},
\end{cases}
\end{equation}

where $A$, $F_d$, $t_{\rm p}$, $\tau_{\rm r}$, and $\tau_{\rm d}$ denote the peak flux, the late-time constant flux level, the time of the flux peak relative to the initial EP/WXT detection, and the rise and decay timescales, respectively. The best-fit parameters are $A = (4.75 \pm 0.08)\times10^{-10}\,\mathrm{erg\,cm^{-2}\,s^{-1}}$, $F_d = (0.68\pm0.02)\times10^{-10}\,\mathrm{erg\,cm^{-2}\,s^{-1}}$, $t_{\rm p}=2.6\pm0.2$~days, $\tau_{\rm r}=2.5^{+0.5}_{-0.4}$~days, and $\tau_{\rm d}=15.7^{+0.5}_{-0.7}$~days.

We note that the EP/WXT flux measurements appear systematically offset from those obtained with other telescopes during the decay phase, likely due to the narrower energy band of EP/WXT.

\subsection{Binary system parameters}
\label{binary}
Assuming a $10\,M_{\odot}$ black hole accretor and adopting a viscosity parameter of $\nu = 10^{15}\,\mathrm{cm^2\,s^{-1}}$~\citep{2007MNRAS.374..466P}, the outer disk radius can be estimated from the exponential decay timescale as

\begin{equation}
R_{\rm disc}(\tau_d)
=
\sqrt{3\nu\tau_d}
\approx
6\times10^{10}\,\mathrm{cm}
\approx
4\times10^4\,R_g .
\end{equation}

Assuming that this radius is comparable to the circularization radius \(R_{\rm circ}\) and a binary mass ratio of $q=0.1$, the orbital period can be estimated using Eq.~4.18 of \citet{1992apa..book.....F}:

\begin{equation}
\begin{aligned}
P_{\rm orb}
=&\,
\frac{2\pi R_\odot^{3/2}}
{\sqrt{G M_\odot}}
\left(
\frac{M_1}{M_\odot}
\right)^{-1/2}
\\
&\times
\left(
\frac{R_{\rm circ}}{R_\odot}
\right)^{3/2}
(1+q)^{-2}
\left[
0.500 - 0.227 \log_{10} q
\right]^{-6} .
\end{aligned}
\end{equation}

This gives an estimated orbital period of $P_{\rm orb}\approx4.3\,\mathrm{hr}$. However, due to observational gaps caused by Sun-avoidance constraints, we cannot exclude the possibility that the source experienced a reflare between 2025 October and 2026 March, which could introduce uncertainties into the system parameter estimations.

\section{Supplementary figures and tables}

\begin{table*}[htbp]
  \centering
  \caption{Log of the X-ray follow-up observations.}
  \label{tab:obs_info}
  \begin{tabular}{l c c c c c c c c c} 
    \hline
    Start Time (UTC) & Instrument & Mode & ObsID & Exposure Time (s) \\
    \hline
    2025-09-17T08:57:58.87 & SVOM/MXT & \dots & 1140857003 & 7074 \\
    2025-09-17T11:34:24.89 & EP/FXT & FXT-A: FF; FXT-B: PW & 08500000408 & 3127 \\
    2025-09-17T21:38:40.10 & Swift/XRT & PC & 03000100001 & 1254 \\
    2025-09-17T23:14:02.46& Swift/XRT & WT & 03000100002 & 1849      \\
    2025-09-18T14:04:33.83 & SVOM/MXT & \dots& 1140857021 & 4727 \\
    2025-09-19T06:14:55.31 & SVOM/MXT & \dots& 1140857023 & 6559 \\ 
    2025-09-19T15:36:09.57& Swift/XRT& WT & 03000100003 & 998      \\
    2025-09-20T06:30:28.03 & SVOM/MXT & \dots & 1140857026 & 11846 \\
    2025-09-20T10:27:26.40& Swift/XRT& WT & 03000100004 & 1638      \\
    2025-09-21T06:46:03.71 & SVOM/MXT & \dots & 1140857027 & 11855 \\
    2025-09-22T15:15:57.36& Swift/XRT& WT & 03000100005 & 784      \\
    2025-09-23T09:45:06.00& Swift/XRT& WT & 03000100006 & 1075      \\
    2025-09-24T11:26:09.00     & NuSTAR       & \dots& 91101336002 & 20706   \\
    2025-09-26T21:38:24.43& Swift/XRT&  WT& 03000100007 & 518     \\
    2025-10-01T02:42:43.38& Swift/XRT&  WT& 03000100008 & 1185      \\
    2025-10-02T04:58:48.06& Swift/XRT&  WT& 03000100009 & 853      \\
    2025-10-03T07:22:02.96& Swift/XRT&  WT& 03000100011 & 942      \\
    2025-10-04T02:03:26.87& Swift/XRT& PC & 03000100010 & 632      \\
    2025-10-05T23:05:37.37& Swift/XRT& PC & 03000100012 & 862      \\
    2025-10-06T00:51:56.20& Swift/XRT& WT& 03000100013 & 907      \\
    2025-10-07T04:32:07.57& Swift/XRT & WT& 03000100014& 716      \\
    2025-10-08T00:47:33.13& Swift/XRT& PC& 03000100015 & 867      \\
    2025-10-11T00:04:24.96& Swift/XRT& PC& 03000100016 & 732      \\
    2025-10-14T01:17:35.47& Swift/XRT& PC& 03000100017 & 924      \\
    2025-10-17T06:38:09.47& Swift/XRT& PC& 03000100019 & 764      \\
    2025-10-18T09:00:13.33& Swift/XRT& PC& 03000100018 & 824      \\
    2025-10-23T16:22:59.63& Swift/XRT& PC& 03000100021 & 839      \\
    2025-11-01T06:33:13.37& Swift/XRT& PC& 03000100022 & 827      \\
    2026-04-22T19:05:30.83 & EP/FXT & FXT-A: FF; FXT-B: PW & 11900697344 & 3659 \\

    \hline
  \end{tabular}
\end{table*}

\begin{table}[htbp]
\centering
\caption{Best-fitting spectral parameters of the joint Swift/XRT and NuSTAR spectra.}
\begin{tabular}{lcc}
\hline
Component & Parameter & Value \\
\hline
Constant & factor (FPMA) &  $1$ (frozen) \\
         & factor (FPMB) & $1.024 \pm 0.007$ \\
         & factor ($\rm XRT_{06}$) &  $0.95_{-0.03}^{+0.02}$   \\
         & factor ($\rm XRT_{07}$) &  $0.91\pm0.03$   \\
\hline
TBabs & $N_{\rm H}$ $(10^{22}\,\mathrm{cm^{-2}})$ &  $0.90 \pm 0.05$ \\
\hline
Cutoffpl & $\Gamma$      &  $2.08_{-0.03}^{+0.02}$  \\
         & $E_{\rm cut}$ (keV)&  $217_{-50}^{+72}$ \\
         & norm  $(\rm photons/keV/cm^2)$        &  $0.157_{-0.007}^{+0.004}$ \\
\hline 
Relxill  & $q_1$        &  $3$ (frozen)       \\
         & $q_2$        &  $3$ (frozen)       \\
         & $R_{\rm br}$~$(R_g)$          &  $1000$ (frozen)    \\
         & a             &  $0.998$ (frozen)   \\
         & Inclination ($\degr$) &  $58_{-31}^{+16}$ \\
         & $R_{\rm in}$~$(R_g)$        &  $100$ (frozen)              \\
         & $R_{\rm out}$~$(R_g)$          &  $1000$ (frozen)    \\
         & z             &  $0$ (frozen)       \\
         & $\Gamma$ &  $=\Gamma_{\rm cutoffpl}$                     \\
         & $\rm log\xi$         &   $1.70_{-0.35}^{+0.09}$   \\
         & $A_{\rm Fe}$           &   $<0.6$        \\
         & $E_{\rm cut}$ (keV) &  $=E_{\rm cutoffpl}$                   \\
         & Reflection fraction &  $-1$ (frozen)         \\
         & norm ($10^{-4}$)  &    $2.1_{-1.0}^{+0.6}$    \\
\hline 
Diskbb   & $T_{\rm in}$ (keV)     &  $0.083 \pm 0.004$  \\
         & norm $(10^7)$ [$ (R_{\rm in}/D_{10})^2 \cos\theta $] &  $3.1_{-1.5}^{+1.9}$     \\
\hline
         & $\chi^2/\rm{d.o.f}$ &    $1853/1662$       \\
\hline
\end{tabular}
\tablefoot{
The spectra are jointly fitted with the model constant*tbabs*(cutoffpl+relxill+diskbb). Errors are given at the 90\% confidence level. $XRT_{06}$ and $XRT_{07}$ correspond to Swift/XRT ObsIDs 03000100006 and 03000100007.}
\label{tab_fit_spec}
\end{table}

\begin{figure}[htbp]
    \centering
    \includegraphics[width=0.75\linewidth]{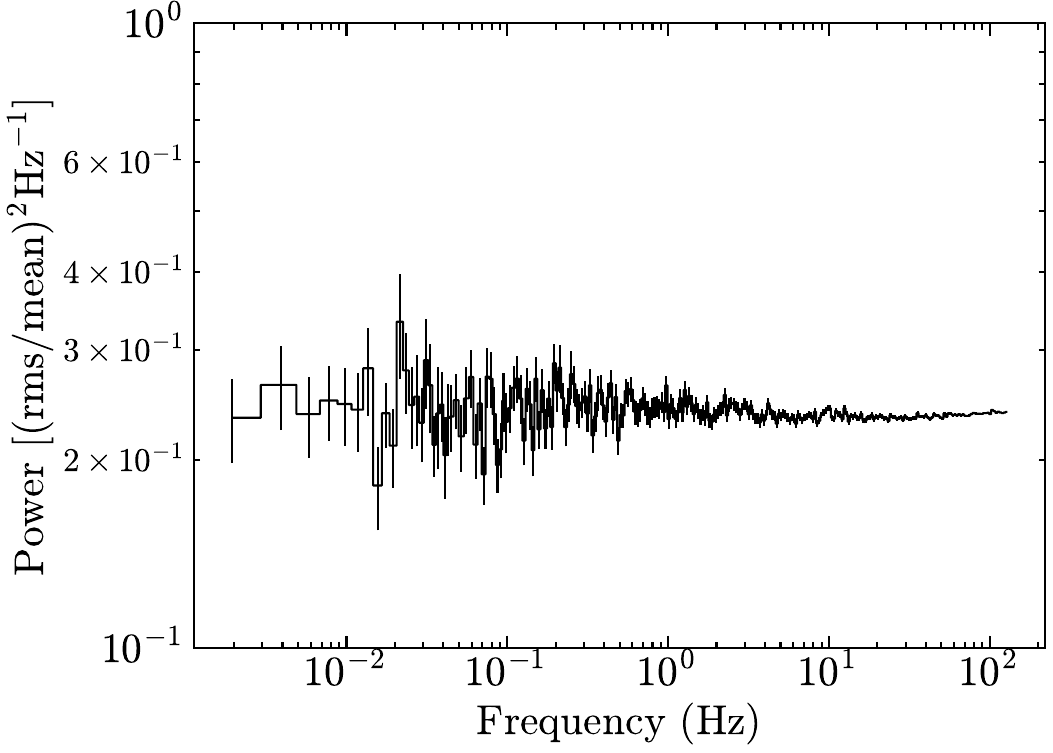}  
    \caption{Averaged power density spectrum of the NuSTAR observation on 2025 September 24 in the 4--78~keV energy band.}
    \label{pds_nustar}
\end{figure}

\begin{figure}
    \centering
    \includegraphics[width=0.75\linewidth]{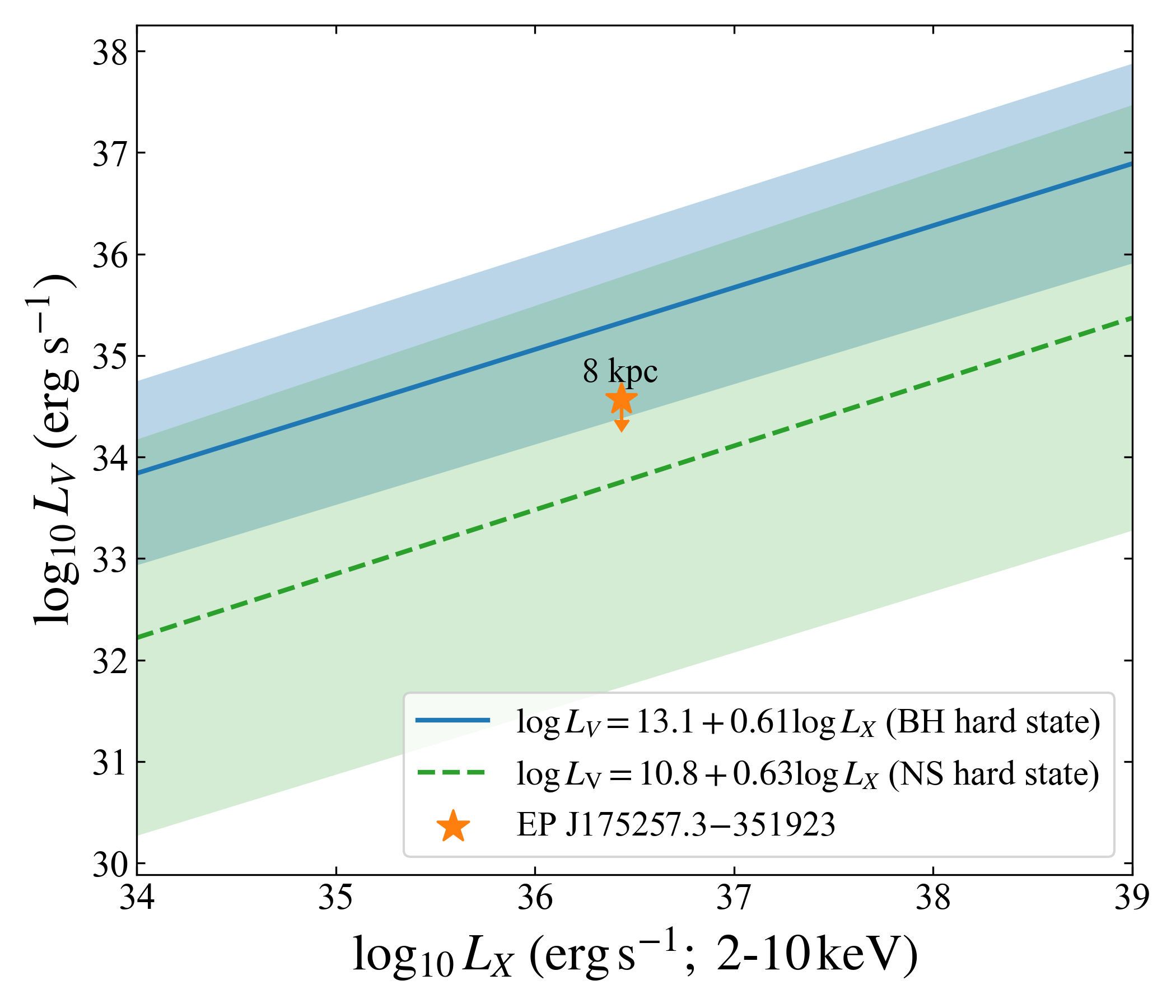}
    \caption{Optical/X-ray luminosity relation for low-mass XRBs. The blue line and shaded region show the empirical correlation and its $1\sigma$ uncertainty for hard-state black-hole XRBs, while the green line and shaded region show those for hard-state neutron-star XRBs \citep{2006MNRAS.371.1334R}. The star symbol marks EP~J175257.3--351923 at an assumed distance of 8~kpc, with a dereddened $V$-band luminosity near the outburst peak of $\lesssim 4\times10^{34}\ {\rm erg\ s^{-1}}$ and a quasi-simultaneous 2--10~keV X-ray luminosity of $\sim3 \times 10^{36}\ {\rm erg\ s^{-1}}$.}
    \label{fig_lv_lx}
\end{figure}

\begin{figure}
    \centering
    \includegraphics[width=0.9\linewidth]{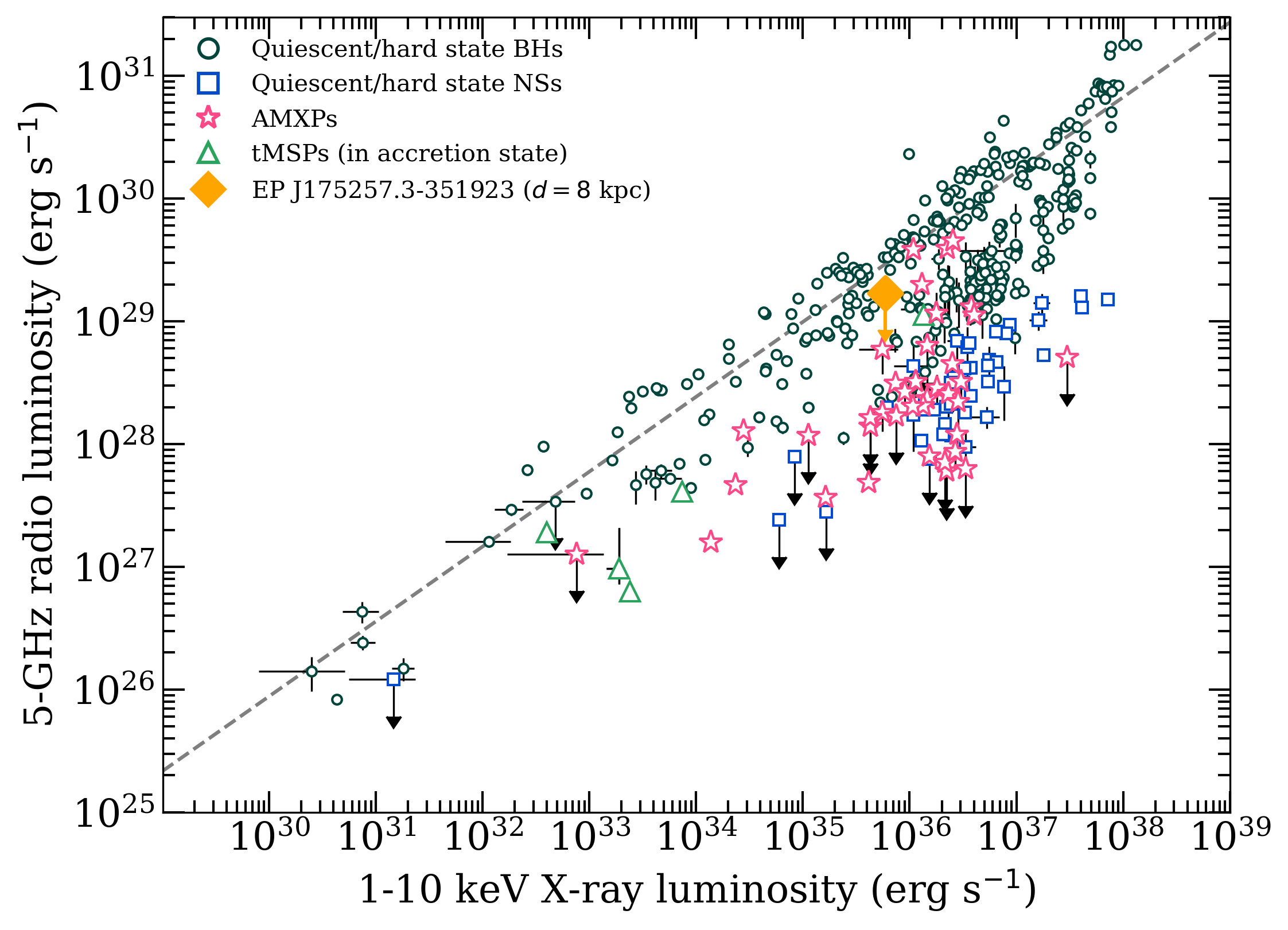}
    \caption{Radio versus X-ray luminosities of low-mass XRBs in the hard state (adapted from \citealt{arash_bahramian_2022_7059313}), with EP~J175257.3--351923 overlaid. The radio constraint is a $3\sigma$ VLASS upper limit at 3~GHz (2026 February~1), converted to 5~GHz assuming a flat radio spectrum. The X-ray luminosity is from the EP/FXT observation on 2026 April~22. Although the X-ray and radio observations are not simultaneous, the source had entered a plateau phase, and the X-ray luminosity is therefore assumed to be comparable at the VLASS epoch. Both luminosities assume a distance of 8~kpc.}
    \label{fig_lrlx}
\end{figure}
\end{appendix}

\end{document}